\documentclass[12pt]{article}

\usepackage{tocloft}
\usepackage{graphicx}
\graphicspath{ {./Plots/} }
\usepackage{natbib}
\usepackage[protrusion=true,expansion=true]{microtype}
\usepackage{amsmath,amsthm,amsfonts,amssymb,mathrsfs,dsfont,mathtools}
\usepackage{booktabs}
\usepackage{dutchcal}
\usepackage{url}
\usepackage{float}
\usepackage{color}
\usepackage{parskip}
\usepackage[explicit]{titlesec}
\usepackage{threeparttable}
\usepackage{setspace}
\usepackage[margin=0.80in]{geometry}
\usepackage[font = onehalfspacing]{caption}
\usepackage[usenames,dvipsnames]{xcolor}
\usepackage[hyperfootnotes=true]{hyperref}
\usepackage[labelfont=bf]{caption}
\usepackage{authblk}
\usepackage{caption}
\usepackage[titletoc,toc,title]{appendix}
\usepackage{enumerate}
\usepackage{soul}
\usepackage{subfig}
\usepackage{accents}
\usepackage{subfiles} 
\usepackage[capitalise, nameinlink]{cleveref}
\usepackage{dsfont}
\usepackage{comment}

\usepackage{changepage}
\usepackage[noend]{algpseudocode}
\usepackage{tocloft}
\usepackage[protrusion=true,expansion=true]{microtype}
\usepackage{amsmath,amsthm,amsfonts,amssymb,mathrsfs,dsfont,mathtools}
\usepackage{float}
\usepackage{color}
\usepackage{parskip}
\usepackage[explicit]{titlesec}
\usepackage{threeparttable}
\usepackage{setspace}
\usepackage[font = onehalfspacing]{caption}
\usepackage[usenames,dvipsnames]{xcolor}
\usepackage[hyperfootnotes=true]{hyperref}
\usepackage[labelfont=bf]{caption}
\usepackage{authblk}
\usepackage[titletoc,toc,title]{appendix}
\usepackage{enumerate}
\usepackage{soul}
\usepackage{subfig}
\usepackage{accents}
\usepackage[capitalise, nameinlink]{cleveref}
\usepackage{diagbox}
\usepackage{array,multirow}

\usepackage[utf8]{inputenc} 
\usepackage[T1]{fontenc}    
\usepackage{hyperref}       
\usepackage{url}            
\usepackage{booktabs}       
\usepackage{amsfonts}       
\usepackage{nicefrac}       
\usepackage{microtype}      
\usepackage{lipsum}		
\usepackage{graphicx}
\usepackage{natbib}
\usepackage{doi}
\usepackage{subfiles}
\usepackage{bbm}

\usepackage{adjustbox}
\usepackage{multirow}
\usepackage{amsmath}
\usepackage[section]{placeins}
\usepackage{algorithm}
\usepackage{algpseudocode}
\usepackage{dsfont}

\DeclareUnicodeCharacter{2212}{-}

\setstcolor{red} 

\hypersetup{colorlinks = true, citecolor = MidnightBlue, linkcolor = MidnightBlue, urlcolor = MidnightBlue}

\titleformat{\section}{\normalfont\large\bfseries}{\thesection}{1em}{#1}
\titleformat{\subsection}{\normalfont\normalsize\bfseries}{\thesubsection}{1em}{#1}
\titleformat{\subsubsection}{\normalfont\normalsize\itshape}{\thesubsubsection}{1em}{#1}
\titlespacing\section{0pt}{12pt plus 4pt minus 2pt}{6pt plus 2pt minus 2pt}
\titlespacing\subsection{0pt}{12pt plus 4pt minus 2pt}{3pt plus 2pt minus 3pt}
\titlespacing\subsubsection{0pt}{12pt plus 4pt minus 2pt}{0pt plus 2pt minus 3pt}

\def\boxit#1{\vbox{\hrule\hbox{\vrule\kern6pt
          \vbox{\kern6pt#1\kern6pt}\kern6pt\vrule}\hrule}}

\definecolor{orange}{rgb}{1,0.5,0}
\definecolor{MyDarkBlue}{rgb}{0,0.08,0.45}

\def\boxit#1{\vbox{\hrule\hbox{\vrule\kern6pt
          \vbox{\kern6pt#1\kern6pt}\kern6pt\vrule}\hrule}}

\definecolor{orange}{rgb}{1,0.5,0}
\definecolor{MyDarkBlue}{rgb}{0,0.08,0.45}

\begin{document}
\title{\Large \bfseries Enhancing Deep Hedging of Options with Implied Volatility Surface Feedback Information\thanks{\footnotesize  We thank John Hull and Zissis Poulos for valuable feedback about this work. We also gratefully acknowledge fruitful discussions with the participants from the Banff International Research Station (BIRS), the CANSSI Ontario Statistics Cast Seminars, Wilfrid Laurier University, and the International Association for Quantitative Finance (IAQF), New York, seminars. Fran\c cois acknowledges a fellowship from the Canadian Derivatives Institute. Gauthier is supported by the Natural Sciences and Engineering Research Council of Canada (NSERC, RGPIN-2019-04029), a professorship funded by HEC Montr\'eal, and the HEC Montr\'eal Foundation. Godin is funded by NSERC (RGPIN-2024-04593). \textbf{Data availability statement}: The raw data (option prices) are available from OptionMetrics. The estimated model parameters are included in the code available at \href{https://github.com/cpmendoza/DeepHedging_JIVR.git}{the github platform}. \textbf{Conflict of interest statement}: No conflict of interest is reported by the authors.}} 

\author[a]{Pascal Fran\c cois}
\author[b]{Genevi\`eve Gauthier}
\author[c]{Fr\'ed\'eric Godin\thanks{Corresponding author.
{ \it Email addresses:} 
\href{mailto:pascal.francois@hec.ca}{pascal.francois@hec.ca} (François),
\href{mailto:genevieve.gauthier@hec.ca}{genevieve.gauthier@hec.ca} (Gauthier), \href{mailto:frederic.godin@concordia.ca}{frederic.godin@concordia.ca} (Godin), \href{mailto:carlos.octavio.perez92@gmail.com}{carlos.octavio.perez92@gmail.com} (P\'erez-Mendoza).}} 
\author[c]{Carlos Octavio P\'erez-Mendoza}

\affil[a]{{\small Department of Finance, HEC Montr\'eal, Canada}}
\affil[b]{{\small GERAD and Department of Decision Sciences, HEC Montr\'eal, Canada}}
\affil[c]{{\small Concordia University, Department of Mathematics and Statistics, Canada}}

\vspace{-30pt}
\date{ \today}


\maketitle \thispagestyle{empty} 

%

\vspace{-20pt}

\begin{abstract}

{\footnotesize We present a dynamic hedging scheme for S\&P 500 options, where rebalancing decisions are enhanced by integrating information about the implied volatility surface dynamics. The optimal hedging strategy is obtained through a deep policy gradient-type reinforcement learning algorithm. The favorable inclusion of forward-looking information embedded in the volatility surface allows our procedure to outperform several conventional benchmarks such as practitioner and smiled-implied delta hedging procedures, both in simulation and backtesting experiments. The outperformance is more pronounced in the presence of transaction costs.}

\noindent \textbf{JEL classification:} C45, C61, G32.


\noindent \textbf{Keywords:} Deep reinforcement learning, optimal hedging, implied volatility surfaces.
\end{abstract}

\medskip

\thispagestyle{empty} \vfill \pagebreak

\setcounter{page}{1}
\pagenumbering{roman}

\doublespacing

\setcounter{page}{1}
\pagenumbering{arabic}


\section{Introduction}\label{se:intro}

Since the advent of the \cite{BlackScholes} framework, dynamic hedging has become a standard financial risk management tool for managing the risk associated with options portfolios. The \cite{BlackScholes} framework has the remarkable property that delta hedging --a hedging strategy invested exclusively in the underlying asset and the money market account-- achieves the perfect replication of a European-style contingent claim. In practice, this property is lost due to frictions. Most notably, considering the infrequent rebalancing of the hedging portfolio, classic delta hedging, which is inherently local, can no longer protect against infinitesimal shocks in the underlying asset price. Such an imperfect hedge inevitably yields a hedging error that has to be managed. 

Several works have subsequently extended the idealized setting of the Black-Scholes framework to account for imperfect hedging, incorporating features such as discrete-time rebalancing \citep{BOYLE1980259}, transaction costs \citep{Leland, boyle1992option, toft1996mean, meindl2008dynamic, zakamouline2009best, lai2009option}, trading constraints \citep{edirisinghe1993optimal} or liquidity costs \citep{frey1998perfect,cetin2010option,gueant2017option}. Another very important avenue for the development of a hedging procedure is the computation of the delta (and other "Greek" sensitivity parameters) which purely relies on market data and does not require stringent postulates about stochastic dynamics of associated risk factors: see among others, \cite{BATES2005195}, \cite{ALEXANDER2007}, and \cite{franccois2021smile}. 
This approach utilizes implied volatility (IV) surfaces to derive the Greeks, specifically the option delta and gamma, based on a mild scale-invariance assumption for the underlying asset return distribution. 

This paper builds on existing literature by developing an approach that incorporates IV surface information to derive optimal hedging positions. Unlike traditional methods focusing on local conditions, we adopt a global perspective, minimizing the risk metric associated with the terminal hedging error in a multi-period horizon framework.
We leverage developments in risk-aware reinforcement learning (RL) to find the sequence of hedge ratios optimizing the hedger's total risk until the option expiry, a setup analogous to \cite{buehler2019deep}'s deep hedging approach. 
The novelty of our work lies in integrating factors that influence IV surface dynamics into the state variable set for determining hedging positions. The hedging agent therefore relies on market expectations for the underlying return distributions over several temporal horizons, producing genuine optimal forward-looking multi-stage hedging decisions informed by IV surfaces. To conduct our numerical experiments, we use the JIVR model of \cite{Frnacois2023}, which is a parsimonious and tractable econometric model representing the joint dynamics of the underlying asset return and five interpretable factors driving the IV surface of the S\&P 500. Notably, the model has been estimated on a data period spanning more than 25 years of daily option data, reflecting market behavior in a wide array of scenarios, including several financial crises.

Hedging procedures with risk-aware reinforcement learning procedures have received substantial attention from the literature recently, see for instance the following non-exhaustive list: \cite{halperin2019qlbs}, \cite{cao2020deep}, \cite{du2020deep}, \cite{carbonneau2021equal}, \cite{carbonneau2021deep}, \cite{giurca2021delta}, \cite{horvath2021deep}, \cite{imaki2021no}, \cite{lutkebohmert2022robust}, \cite{xu2022delta}, \cite{cao2023gamma}, \cite{carbonneau2023deep}, \cite{Marzaban_rl}, \cite{mikkila2023empirical}, \cite{pickard2023deep}, \cite{raj2023quantum}, \cite{wu2023robust} and \cite{neagu2024deep}. Nevertheless, to the best of our knowledge, our paper is the first to conduct multi-stage RL-based hedging that incorporates IV surface information directly as state variables.

This study focuses on hedging procedures using the underlying asset as the only hedging instrument. Hedging strategies that add other instruments such as options potentially improve the quality of the hedge (by managing other Greeks like the gamma or the vega) but they come with higher transaction costs. In addition, they require a precise timely execution of trades between the option and the underlying asset markets. As a matter of fact, delta hedging has been a long-lasting risk management problem that continues to draw the attention of researchers \citep{hull2017optimal}. It remains relevant for several market participants including option market makers \citep{huh2015options, kokholm2024model}, although the extent of their delta hedging has been recently challenged \citep{hu2023options}, insurance companies \citep{augustyniak2012out, sun2019benchmarked}, structured products issuers \citep{entrop2020hedging, henderson2020pre}, or proprietary trading firms \citep{ni2005stock, o2023retail}.

Our RL approach is shown to substantially outperform delta-based hedging strategies acting as benchmarks. In particular, RL agents trained with asymmetric objective functions such as the conditional value-at-risk (CVaR) or the semi-mean squared-error (SMSE) offer superior tradeoffs between profitability and downside risk. 
The outperformance of RL agents is even more pronounced in the presence of transaction costs, as such agents manage to develop hedging strategies that remain efficient in terms of risk mitigation while generating lower turnover. 
A variable importance analysis highlights that the conditional variance of the underlying asset returns, the level of the IV surface and its slope all significantly influence hedging performance, irrespective of the risk metric employed within the objective function.

The out-of-sample backtesting study highglights that RL-based hedging strategies can outperform traditional benchmarks, particularly when the algorithm is given access to rich market information such as the implied volatility (IV) surface. In the presence of transaction costs, RL approaches demonstrate superior hedging performance. However, in the absence of transaction costs, the benefits of RL hedging are contingent on the quality of the informational inputs—without IV surface data, the RL agent fails to outperform delta hedging in terms of mean squared error. These findings highlight the potential of reinforcement learning as a flexible and powerful tool for risk management, provided that it is informed with relevant market signals.

The rest of the paper is organized as follows. \cref{se:hedgepro} frames the hedging problem in terms of a deep reinforcement learning framework. \cref{se:JMD} provides the components of the JIVR model. \cref{se:SSG} presents the numerical results, assessments, and global feature importance analysis.\footnote{The Python code to replicate the numerical experiments from this paper can be found at the following link: \href{https://github.com/cpmendoza/DeepHedging_JIVR.git}{https://github.com/cpmendoza/DeepHedging\_JIVR.git}.} \cref{subsub:benchmarking_realpaths} presents the out-of-sample backtest study.  \cref{se:conclusion} concludes.


\section{The hedging problem}\label{se:hedgepro}

The mathematical formulation of the hedging problem considered herein, along with the solution approach based on deep reinforcement learning, are described in this section. 

\subsection{The hedging optimization problem}\label{subse:hedgepro_formulation}

This paper introduces a dynamic hedging strategy for European-style options that leverages insights provided by the implied volatility surface. The approach aims to minimize some risk measure applied to the terminal hedging error.

The European option payoff $\Psi(S_{T})$ depends on the price of the underlying asset at maturity, denoted as $T$ trading days. The hedging strategy involves managing a self-financing portfolio composed of both the underlying asset and a risk-free asset, with daily rebalancing.  The strategy is represented by the predictable process 
$\{ (\phi_{t}, \delta_{t}) \}_{t=1}^{T}$, where $\phi_{t}$ is the cash held at time $t-1$ and carried forward to the next period, and $\delta_{t}$ denotes the number of shares of the risky asset $S$ held during the interval $(t-1, t]$.
The time-$t$ portfolio value is
\begin{equation*} 
    V_{t}^{\delta}=\phi_{t}\mbox{e}^{r_{t}\Delta}+\delta_{t} S_{t}\mbox{e}^{q_{t}\Delta} 
\end{equation*}
where $r_{t}$ is the time-$t$ annualized continuously compounded risk-free rate, $q_{t}$ is the annualized underlying asset dividend yield, both on the interval $(t-1,t]$, and $\Delta=\frac{1}{252}$.
To account for transaction costs the self-financing condition entails 
\begin{equation}
\label{self-financing}
    \phi_{t+1} +\delta_{t+1}S_{t}= V_{t}^{\delta}-\kappa S_{t}\mid \delta_{t+1} - \delta_{t}\mid ,
\end{equation}
where $\kappa$ is the rate of proportional transaction costs.

The optimal hedging problem is an optimization task where an agent seeks to minimize the risk exposure associated with a short position in the option. More precisely, it is a sequential decision problem where the agent looks for the best sequence of actions $\delta= \{ \delta_t\}^T_{t=1}$ that minimizes a penalty function $\rho$ applied to the hedging error at maturity for a short position, defined as 
\begin{equation*}
    \xi_{T}^{\delta} = \Psi(S_{T}) - V_{T}^{\delta}.
\end{equation*}
Note that $\xi_{T}^{\delta}$ is a loss variable, with profits being represented by $-\xi_{T}^{\delta}$.
The problem is
\begin{equation}\label{Hedging_problem}
\delta^{*}=\mathop{\arg \min}\limits_{\delta\ } \rho  \left( \xi_{T}^{\delta} \right) ,
\end{equation}
where $\rho$ is a risk measure, acting as the penalty function. Each time-$t$ action $(\phi_{t+1},\delta_{t+1})$ is of feedback-type, with such decision being a function of current available information on the market: $\delta_{t+1}= \tilde{\delta}(X_t)$ for some function $\tilde{\delta}$ with state variables vector $X_t$. 
Section \ref{subsub:state_space} further describes these state variables that include the underlying asset current value as well as some information about the implied volatility surface, among others. Due to Equation \eqref{self-financing}, $\phi_{t+1}$ is fully characterized when $\delta_{t+1}$ is specified, and as such the time-$t$ action to be chosen is simply $\delta_{t+1}$.

In this paper we consider three penalty functions that are very popular in the literature:
\begin{itemize}
    \item Mean Square Error (MSE): $\rho(\xi_{T}^{\delta})= \mathbb{E}[(\xi_{T}^{\delta})^{2}]$.
    \item Semi Mean-Square Error (SMSE): $\rho(\xi_{T}^{\delta})= \mathbb{E}[(\xi_{T}^{\delta})^{2}\mathbbm{1}_{\{\xi_{T}^{\delta}\geq 0\}}]$.
    \item Conditional Value-at-Risk (CVaR$_\alpha$): $\rho(\xi_{T}^{\delta})=\mathbb{E}[\xi_{T}^{\delta}\mid \xi_{T}^{\delta}\geq \mbox{VaR}_{\alpha}(\xi_{T}^{\delta})]$, where $\alpha \in (0,1)$ and $\mbox{VaR}_{\alpha}(\xi_{T}^{\delta})$ is the Value-at-Risk defined as $\mbox{VaR}_{\alpha}(\xi_{T}^{\delta}) = \min_{c}\{ c:\mathbb{P}(\xi_{T}^{\delta}\leq c)\geq \alpha \}$ and $\alpha=95\%$ or $99\%$ in this work.
\end{itemize}
The MSE was first proposed in the seminal variance-optimal hedging framework of \cite{Schweizer}, which was later extended to the multivariate case by \cite{Remillard}. The SMSE is a particular case of the asymetric polynomial penalty considered for instance subsequently in \cite{pham2000dynamic}, \cite{FRANCOIS2014312} and \cite{carbonneau2023deep}. It provides the advantage over the MSE to avoid penalizing hedging gains. Lastly, we consider CVaR as a standard metric for measuring potential catastrophic tail events, frequently mandated by financial regulators for use in financial institutions. This metric has also been explored in global hedging contexts in \cite{melnikov2012dynamic}, \cite{GodinCVaR}, \cite{buehler2019deep}, \cite{carbonneau2021equal} or \cite{cao2023gamma}, among others. 

Historically, an alternative method for addressing the problem described in Equation \eqref{Hedging_problem} employs backward recursion within dynamic programming frameworks. However, this approach is hindered by the curse of dimensionality, limiting its practical applicability. 
To overcome these limitations, we tackle Problem \eqref{Hedging_problem} using reinforcement learning.

\subsection{Reinforcement learning and deep hedging}\label{subsec:state_space}

The optimal hedging problem \eqref{Hedging_problem} can be formulated as a reinforcement learning (RL) problem because it is a feedback sequential decision-making task. In this framework, an agent learns a policy (the investment strategy $\delta$) which dictates trading actions to be applied as a function of state variables to minimize the hedging objective highlighted in Equation \eqref{Hedging_problem}. More precisely, the problem consists in learning the mapping $\tilde{\delta}$.

Consistently with \cite{buehler2019deep}, the problem established in \eqref{Hedging_problem} is solved by direct estimation of the policy through a deep policy gradient approach that relies on the estimation of the mapping $\tilde{\delta}$ by an Artificial Neural Network (ANN). Denoting by $\tilde{\delta}_{\theta}$ the policy obtained when $\tilde{\delta}$ is estimated by an ANN with parameters $\theta$, the objective function we need to minimize is thus
\begin{equation}\label{penaltyfunction}
    \mathcal{O}(\theta) = \rho \left( \xi_{T}^{\tilde{\delta}_{\theta}} \right)\mbox{.}
\end{equation}

\subsubsection{Neural network architecture}
\label{se:NNARCHITECTURE}

We propose an architecture reminiscent of these used in \cite{koochali2019probabilistic} and \cite{fecamp2020deep}, which consists in a Recurrent Neural Network with a Feedforward Connection (RNN-FNN) that combines the traditional Long Short-Term Memory Network (LSTM) and the Feedforward Neural Network (FFNN) architectures. The inclusion of LSTM layers mitigates problems related to vanishing gradients.\footnote{Vanishing gradients arise when the gradients of the penalty function become extremely small, leading to slow or halted training (details can be found in \cite{goodfellow2016deep}).} 
The RNN-FNN network is defined as a composition of LSTM cells $\{ C_{l} \}_{l=1}^{L_{1}}$ and FFNN layers $\{ \mathcal{L}_{j} \}_{j=1}^{L_{2}}$ under the following functional representation:
\begin{equation*}
    \tilde{\delta}_{\theta}(X_{t})=(\underbrace{ \mathcal{L}_{J} \circ \mathcal{L}_{L_{2}}\circ \mathcal{L}_{L_{2}-1}\circ ... \circ \mathcal{L}_{1}}_{\mbox{FFNN layers}}\circ \underbrace{C_{L_{1}}\circ C_{L_{1}-1}...\circ C_{1}}_{\mbox{LSTM cells}})(X_{t}).
\end{equation*}

 The LSTM cell $C_{l}$ is a mapping that transforms a vector $Z^{(C,\,l-1)}_{t}$ of dimension $d^{(C,\, l-1)}$ into a vector $Z^{(C,\, l)}_{t}$ of dimension $d^{(C,\, l)}$ based on the following equations, considering $Z^{(C,\, 0)}_{t}=X_{t}$:
\begin{align*}
    i^{(l)} &= \mbox{sigm}(W_{i}^{(l)}Z_{t}^{(C,\, l-1)}+b_{i}^{(l)})\mbox{,}\\
    o^{(l)} &= \mbox{sigm}(W_{o}^{(l)}Z_{t}^{(C,\, l-1)}+b_{o}^{(l)})\mbox{,}\\
    c^{(l)} &= i^{(l)}\odot \mbox{tanh}(W_{c}^{(l)}Z_{t}^{(C,\, l-1)}+b_{c}^{(l)})\mbox{,}\\
    Z_{t}^{(C,\, l)} &= o_{t}^{(l)}\odot \mbox{tanh}(c^{(l)})\mbox{,}
\end{align*}
where $\mbox{sigm}(\cdot)$ and $\mbox{tanh}(\cdot)$ are respectively the sigmoid and hyperbolic tangent functions applied element-wise and $\odot$ is the Hadamard product. Layer $\mathcal{L}_{j}$ represents a FFNN layer that maps the input vector $Z^{(\mathcal{L},\, j-1)}_{t}$ of dimension $d^{(\mathcal{L},\, j-1)}$ into a vector $Z^{(\mathcal{L},\, j)}_{t}$ of dimension $d^{(\mathcal{L},\, j)}$ by applying a linear transformation $T_{\mathcal{L}_{j}}(Z^{(\mathcal{L},\, j-1)}_{t})=W_{\mathcal{L}_{j}}Z^{(\mathcal{L},\, j-1)}_{t} + b_{\mathcal{L}_{j}}$ and, subsequently, an element-wise non-linear activation function $g_{\mathcal{L}_{j}}$, i.e., $\mathcal{L}_{j}(Z^{(\mathcal{L},\, j-1)}_{t})=(g_{\mathcal{L}_{j}} \circ T_{\mathcal{L}_{j}})(Z^{(\mathcal{L},\, j-1)}_{t})$ for $j\in\{1,...,L_{2},J\}$, considering $Z^{(\mathcal{L},\, 0)}_{t}=Z^{(C,\, L_{1})}_{t}$.

The trainable parameters $\theta$ of the RNN-FNN network are listed below:
\begin{itemize}
    \item  If $L_{1}\geq l\geq 1$: $W_{i}^{(l)}$, $W_{o}^{(l)}$, $W_{c}^{(l)}\in \mathbb{R}^{d^{(C,\, l)}\times d^{(C,\, l-1)}}$ and $b_{i}^{(l)}$, $b_{o}^{(l)}$, $b_{c}^{(l)}\in \mathbb{R}^{d^{(C,\, l)}\times 1}$ with $d^{(C,\, 0)}$ defined as the original dimension of the network input.
    \item If $L_{2}\geq j\geq 1$: $W_{\mathcal{L}_{j}}\in \mathbb{R}^{d^{(\mathcal{L},\, j)}\times d^{(\mathcal{L},\, j-1)}}$ and $b_{\mathcal{L}_{j}}\in \mathbb{R}^{d^{(\mathcal{L},\, j)}}$ with $d^{(\mathcal{L},\, 0)}=d^{(C,\, L_{1})}$.
    \item If $j=J$: $W_{\mathcal{L}_{J}}\in \mathbb{R}^{1\times d^{(\mathcal{L},\, L_{2})}}$ and $b_{\mathcal{L}_{J}}\in \mathbb{R}$.
\end{itemize}

The selected hyperparameter values for our experiments are detailed in Section \ref{subsub:network_architecture}.

\subsubsection{Neural network optimization}\label{subse:neural optimization}

The RNN-FNN network is optimized with the Mini-batch Stochastic Gradient Descent method (MSGD). This training procedure relies on updating iteratively all the trainable parameters of the network  based on the recursive equation
\begin{equation}\label{updatingrule}
    \theta_{j+1} = \theta_{j}-\eta_{j}\nabla_{\theta} \hat{\mathcal{O}}(\theta_{j}),
\end{equation}
where $\theta_{j}$ is the set of parameters obtained after iteration $j$, $\eta_{j}$ is the learning rate (step size) which determines the magnitude of change in parameters on each time step,\footnote{This parameter can be either deterministic or adaptive, i.e., it may be adjusted during the training period. For more details on this, please refer to \cite{goodfellow2016deep}.} 
$\nabla_{\theta}$ is the gradient operator with respect to $\theta$ and $\hat{\mathcal{O}}$ is the Monte Carlo estimate of the objective function \eqref{penaltyfunction} computed on a mini-batch. Additional details are provided in Appendix \ref{appen:MSGDtraining} of the supplementary material. 


\section{Joint market dynamics}\label{se:JMD}

This section describes the JIVR model of \cite{Frnacois2023}, which provides market dynamics that we use to represent the joint evolution of the S\&P 500 index price and its associated Implied Volatility (IV) surface. Such model is used to construct the state space of the hedging problem. The JIVR model takes the form of an econometric model directly specifying the evolution of the shape of the volatility surface.\footnote{The direct modeling of the volatility surface is aligned with the so-called "instrumental approach" philosophy detailed in \cite{rebonato2004volatility}, which exhibits more flexibility and treats volatility surfaces as objects possessing intrinsic characteristics on their own, not being fully characterized by the underlying asset.} The model has been crafted through the careful analysis of option price data spanning multiple economic periods and crises, with factor loadings being designed to mimic the most important loadings from a principal component analysis on volatility surfaces. The advantages of the JIVR model are twofold. First, the empirical approach used for its construction, which relies on daily fitting to IV data, enables the accurate representation of all segments on the IV surface on a daily basis. Second, it is very tractable and does not rely on the filtering of latent factors driving volatility dynamics as in multiple competing models that rely on risk-neutral dynamics to construct volatility surfaces.

\subsection{Daily implied volatility surface representation}

On any given day, the cross-section of option prices on the S\&P 500 index is captured through the IV surface model introduced by \cite{Frnacois2022} which characterizes the entire surface parsimoniously with a linear combination of five factors. 
More precisely, the time-$t$ IV of an option with time-to-maturity $\tau_{t}=\frac{T-t}{252}$ years and (scaled) moneyness $M_{t}=\frac{1}{\sqrt{\tau_{t}}}\mbox{log}\frac{S_{t}e^{(r_{t}-q_{t})\tau_{t}}}{K}$, where $K$ is the strike price, is
\begin{align}
     & \sigma (M_{t},\tau_{t},\beta_{t}) =  \sum_{i=1}^{5}\beta_{t,i}f_{i}(M_{t},\tau_{t})
     \label{5factormodel_linear}
\end{align}

where $\beta_{t}=(\beta_{t,1},\beta_{t,2},\beta_{t,3},\beta_{t,4},\beta_{t,5})$ stands for the IV factor coefficients at time $t$ and the functions $\{f_{i} \}_{i=1}^{5}$ represent the long-term at-the-money (ATM) level, the time-to-maturity slope, the moneyness slope, the smile attenuation and the smirk, respectively (see Appendix \ref{subappen:IVsurface} of the supplementary material for their specification).
Following the same data processing and estimation procedure outlined in the aforementioned study, we extract the daily time series of the IV factor coefficients spanning from January 4, 1996, to December 31, 2020.\footnote{The sample comes from OptionMetrics database, to which the conventional data exclusion filters are applied.}

\subsection{Joint Implied Volatility and Return (JIVR)}\label{subsec:JIVR}

The JIVR model proposed by \cite{Frnacois2023} leverages the IV representation \eqref{5factormodel_linear} and provides explicit joint dynamics for the IV surface and the S\&P 500 index price.

The first building block of the JIVR model represents the daily underlying asset excess log-return, $R_{t+1}=\mbox{log} \left(\frac{S_{t+1}}{S_{t}}\right)-(r_{t}-q_{t})\Delta$. It integrates an NGARCH(1,1) process with Normal Inverse Gaussian (NIG) innovations to capture volatility clusters and reproduce fat tailed asymmetric returns, while borrowing information from the volatility surface to anchor the evolution of the conditional variance process $h_{t,R}$. 
The second building block includes a multivariate heteroskedastic autoregressive processes with non-Gaussian innovations for all implied volatility factor coefficients. 
The multivariate time series representation of the JIVR model is presented in detail in \mbox{Appendix \ref{subappen:JIVR_timeseries}} of the supplementary material.

The full model is characterized by the current market conditions $\left(S_t,\{ \beta_{t,i} \}^5_{i=1}, \beta_{t-1,2}, h_{t,R}, \{ h_{t,i}\}^5_{i=1}  \right)$, which are respectively the underlying S\&P 500 index price, IV factor coefficients, and conditional variances for the S\&P 500 return and for such coefficients.

Following \cite{Frnacois2023}, the maximum likelihood estimation is applied to S\&P 500 excess returns alongside the time series estimates of surface coefficients $\{ \beta_{t,i} \}^5_{i=1}$. As a byproduct, we obtain daily estimates of conditional variance series $h_{t,R}$ and $\{ h_{t,i}\}^5_{i=1}$ for the time period extending between January 4, 1996 and December 31, 2020.  


\section{Numerical study}\label{se:SSG}

In this section, simulation and backtesting experiments are conducted to evaluate the performance of the proposed hedging approach.

\subsection{Stochastic market generator}\label{subsec:simulator}

\subsubsection{Market simulator}

The JIVR model is used in subsequent simulation experiments to generate paths of the variables pertaining to market dynamics $(S_{t},\{\beta_{t,i} \}^5_{i=1}, h_{t,R}, \{ h_{t,i}\}^5_{i=1})$, which drive the hedging decisions.\footnote{Coefficient $\beta_{t-1,2}$ is omitted from the state space due to its minor impact.} The initial conditions $\left(\{ \beta_{0,i} \}^5_{i=1}, h_{0,R}, \{ h_{0,i}\}^5_{i=1}  \right)$ are randomly chosen among values prevailing in our data sample extending between January 4, 1996 and December 31, 2020. This constitutes a wide variety of states of the economy. Following the determination of initial values, the simulation progresses through all time steps in two distinct phases. First, a sequence of NIG innovations $\{ \epsilon_{t} \}_{t=1}^{T}$ is simulated using the Monte Carlo method considering the dependence structure of contemporaneous innovations. Second, equations of Section \ref{subappen:JIVR_timeseries} from the supplementary material are used to obtain values for $\left( R_t, \{\beta_{t,i} \}^5_{i=1}, h_{t,R}, \{ h_{t,i}\}^5_{i=1}  \right)$ for $t=1,\ldots,T$ based on the simulated innovations.

\subsubsection{Market parameters for numerical experiments}

The initial underlying asset value is normalized to $S_0=100$ for simplicity. The options being hedged are assumed to be European call options ($\Psi(S_T)=\mbox{max}(S_T-K,0)$) with maturities $T\in \{21,63,126 \}$ days for short-term, medium-term and long-term maturities, and strikes $K\in \{90,100,110 \}$ for in-the-money (ITM), at-the-money (ATM) and out-of-the-money (OTM) options, respectively.\footnote{Unreported numerical results of European put options exhibit a similar pattern due to the Put-Call parity formula.}
Various levels of proportional transaction cost are considered, namely $\kappa \in \{ 0\%,0.05\%,0.5\%,1\% \}$. The initial value of the hedging strategy, $V_{0}^{\delta}$, is set to the price of the option being hedged, which is provided by the prevailing implied volatility surface.
The annualized continuously compounded risk-free rate and dividend yield are assumed to be constant and are estimated using data from the period 2016 to 2020, with values set to $r=2.66\%$ and $q=1.77\%$, respectively.
Other parameters of the model are specified in Section \ref{subappen:JIVR_timeseries}.

\subsection{Benchmarks}\label{subse:benchmarks}

We benchmark RL hedging strategies against the performance of three delta hedging strategies using the following deltas: (i) the practitioner's Black-Scholes delta hedging (DH) which applies the current implied volatility into the Black-Scholes formula to obtain the hedged option's delta; (ii)  the \cite{Leland} delta (DH-L),
which modifies the DH to reflect proportional transaction costs; and (iii) the Smile-implied delta (SI) introduced by \cite{BATES2005195} and whose closed-form expression for the IV model \eqref{5factormodel_linear} is provided by \cite{Frnacois2022}. The explicit formulas for these three benchmarks are detailed in Appendix \ref{appen:benchmarks}.

\subsection{Neural network settings}\label{sbse:fine-tuning}

\subsubsection{Neural network architecture}\label{subsub:network_architecture}

We employ the RNN-FNN architecture from Section \ref{se:NNARCHITECTURE} with two LSTM cells ($L_{1}=2$) of width 56  ($d_{i}=56$ for $i=1,2$) and two FFNN-hidden layers ($L_{2}=2$) of width 56 with ReLU activation function (i.e., $g_{\mathcal{L}_{i}}(x)=\mbox{max}(0,x)$ for $i=1,2$). In the context of the output FFNN layer $\mathcal{L}_{J}$, which maps the output of hidden layers $Z^{(\mathcal{L},\, L_{2})}_{t}$ into the position in the underlying asset $\delta_{t}$, the standard deep hedging framework typically employs a linear activation function. However, this activation function tends to induce RL agents to adopt doubling strategies, increasing the position in the underlying asset several orders of magnitude over the current position on each period after any loss until the cumulative loss amount is fully recovered. Such strategies are definitely undesirable from the perspective of sound risk management. Therefore, we opt to introduce a leverage constraint through the output layer activation function.
This leverage constraint is a dynamic upper bound on the activation function, denoted as $g_{\mathcal{L}_{J}}(Z,t)=\mbox{min}(Z,B_{t}(Z))$, with $Z=T_{\mathcal{L}_{J}}(Z^{(\mathcal{L},\, L_{2})}_{t})$ representing the typical deep hedging underlying asset position, and $B_{t}(Z)$ the highest position in the index
that can be held in the portfolio. 
Such an upper bound avoids excessive leverage and limits the borrowing capacity, i.e., the cash held satisfies $\phi_{t+1}(X_{t})\geq -B$ for all $X_{t}$ and $t$, and $B>0$.\footnote{This type of leverage condition has been previously employed in the literature. For instance, the seminal paper of \mbox{\cite{harrison1981martingales}} assumes a restricted borrowing capacity to maintain a positive wealth throughout the entire hedging period.} The latter, in conjunction with the self-financing constraint \eqref{self-financing}, establishes the dynamic upper bound as
\begin{equation}\label{boundary_condition}
    B_{t}(Z) = \left\{
     \begin{array}{@{}l@{\thinspace}l}
       \frac{V_{0}+B}{S_{0}}  & \text{ if } t=0\\
       \frac{V_{t}+B+\kappa S_{t}\delta_{t}}{S_{t}(1+\kappa)} & \text{ if } t>0 \text{ and } Z\geq \delta_{t} \\
       \frac{V_{t}+B-\kappa S_{t}\delta_{t}}{S_{t}(1-\kappa)} & \text{ if } t>0 \text{ and } Z< \delta_{t}. \\
     \end{array}
   \right.
\end{equation}
Appendix \ref{se:Bounded strategies - leverage constraint} of the supplementary material provides numerical experiments that illustrate the necessity of including a leverage constraint in the network architecture to alleviate the aforementioned issues.

Our numerical experiments demonstrate the superiority of the RNN-FNN architecture over standalone LSTM and FFNN networks, considering a leverage constraint of $B=100$ for all agents. These results are presented in Appendix \ref{appen:architecture_selection} of the supplementary material. Additionally, agents are trained as described in Section \ref{subse:neural optimization} on a training set of 400,000 independent simulated paths with mini-batch size of 1,000 and a learning rate of 0.0005 that is progressively adapted by the ADAM optimization algorithm.\footnote{ADAM is a dynamic learning rate algorithm engineered to accelerate training speeds in deep neural networks and achieve rapid convergence, details can be seen in \cite{goodfellow2016deep}.} In addition, we include a regularization method called dropout with parameter $p=0.5$ to reduce the likelihood of overfitting and enhance the model performance on unseen data.\footnote{Full details of the dropout regularization method can be seen in \cite{goodfellow2016deep}. The selection of $p=0.5$ stems from our numerical experiments detailed in Appendix \ref{appen:dropout_parameter} of the supplementary material, indicating that this value consistently outperforms others across all penalty functions.} The training procedure is implemented in Python, using Tensorflow and considering the \cite{glorot2010understanding} random initialization of the initial parameters of the neural network. Numerical results are obtained from a test set of 100,000 independent paths. 

\subsubsection{State space selection}\label{subsub:state_space}

The nature of the hedging problem, combined with the JIVR model, establishes a decision framework where the optimal decision is entirely defined by the state variables at time $t$, denoted as $X_{t}=(V_{t}^{\delta},\delta_{t},\tau_{t},S_{t},\{\beta_{t,i} \}^5_{i=1}, h_{t,R}, \{ h_{t,i}\}^5_{i=1})$. 

However, implementing numerical methods within the deep policy gradient approach across the entire state space may lead to overfitting. In fact, other studies utilizing the RL framework have reported optimal results with a reduced state space. For instance, \cite{buehler2019deep} and \cite{cao2023gamma} employ the deep hedging algorithm without including the portfolio value $V_{t}^{\delta}$ in the state space configuration. 
Similarly, \cite{carbonneau2021deep} employs the deep hedging framework in a frictionless market without including $\delta_{t}$ in the state space.
We explore three different state space configurations. The first configuration aims to replicate the full state space, denoted by 
\begin{equation}
    (V_{t}^{\delta},\delta_{t},\tau_{t},S_{t},\{\beta_{t,i} \}^5_{i=1}, h_{t,R}, \{ h_{t,i}\}^5_{i=1}), 
\end{equation}
typically considered in a dynamic optimization task with the portfolio value as a state variable. The second state space does not consider the portfolio value and also omits the variance of the IV coefficients $\{ h_{t,i}\}^5_{i=1}$ under the intuition that these coefficients have a second-order effect, with the IV surface coefficients $\{\beta_{t,i} \}^5_{i=1}$ capturing most of the variability. Hence, the reduced state space is defined as
\begin{equation*}
    (\delta_{t},\tau_{t},S_{t},\{\beta_{t,i} \}^5_{i=1}, h_{t,R}). 
\end{equation*}
The current position $\delta_t$ is required for the RL agent to learn about the transaction cost associated with the next rebalancing.  This state space component is no longer useful when $\kappa=0$.  For that reason, we remove this component in absence of transaction cost. In such case, the reduced state space is denoted by 
\begin{equation*}
    (\tau_{t},S_{t},\{\beta_{t,i} \}^5_{i=1}, h_{t,R}). 
\end{equation*}
Our numerical experiments consider the three aforementioned state spaces  to hedge a European ATM call option with maturity $N=63$ days, while taking into account four different transaction cost rates $\kappa \in \{ 0\%, 0.05\%, 0.5\%, 1\% \}$. The impact of the option moneyness and maturity on hedging performance are studied in later subsections. Additionally, our numerical experiments involve four RL agents: RL-CVaR$_{95\%}$, RL-CVaR$_{99\%}$, RL-MSE, and RL-SMSE. These agents aim to minimize different penalty functions.

\begin{figure}[h]\centering
    \caption{Optimal penalty function value for a short position in an ATM call option with maturity of 63 days under various state spaces and transaction cost levels.}
    \includegraphics[width=16cm]{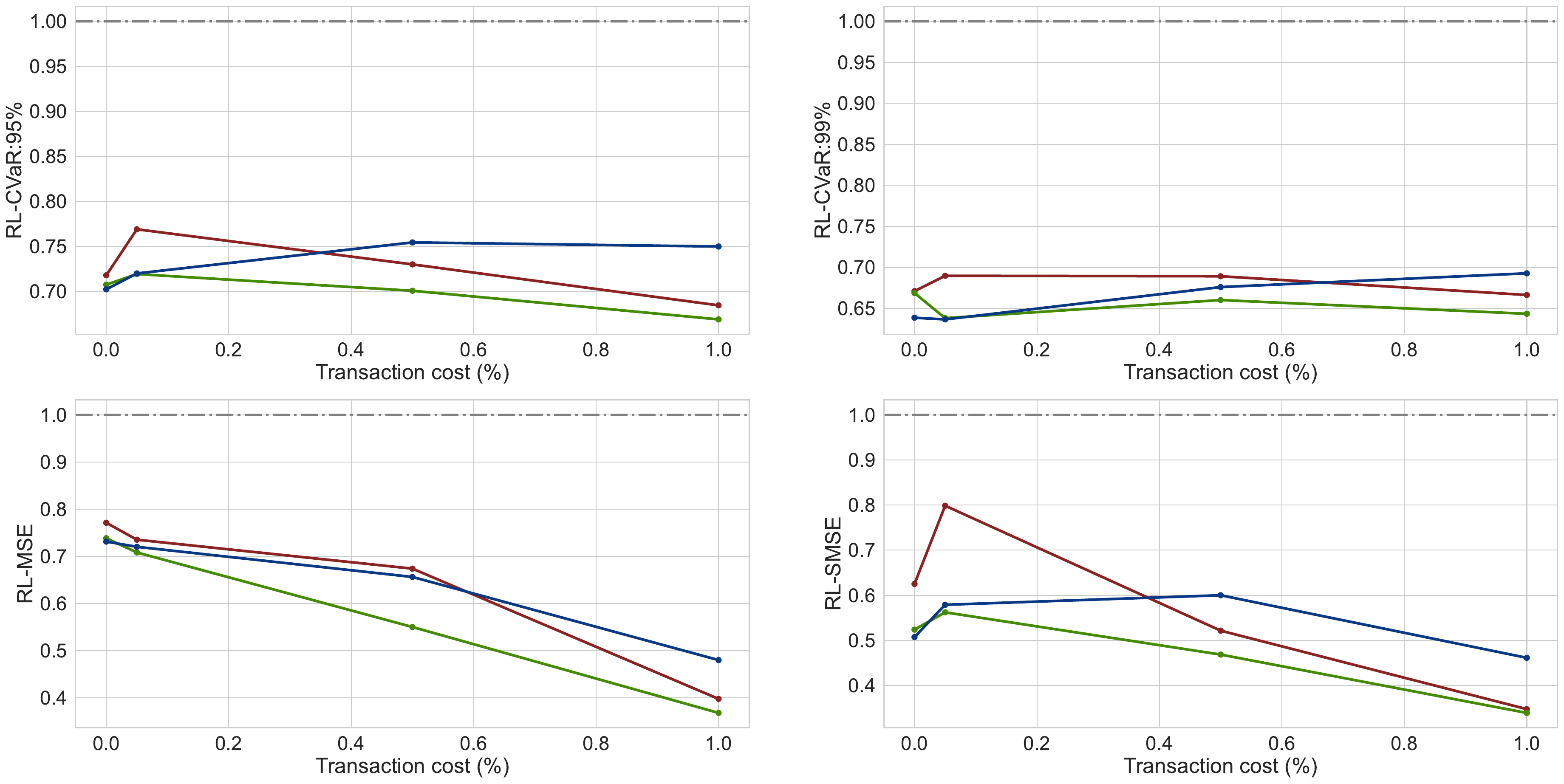}
    \begin{tablenotes}
    \item \footnotesize Results are computed using 100,000 out-of-sample paths according to the conditions outlined in Section \ref{subsub:network_architecture}. Each panel illustrates the optimal penalty function value of an RL agent considering four transaction cost levels. These values are normalized by the estimated values of each penalty function obtained with DH-L for each transaction cost level. Agents are trained under the specified penalty functions considering three state spaces: full state space $(V_{t}^{\delta},\delta_{t},\tau_{t},S_{t},\{\beta_{t,i} \}^5_{i=1}, h_{t,R}, \{ h_{t,i}\}^5_{i=1})$ (red curve), and the two reduced state spaces $(\delta_{t},\tau_{t},S_{t},\{\beta_{t,i} \}^5_{i=1}, h_{t,R}, \{ h_{t,i}\}^5_{i=1})$ (green curve) and $(\tau_{t},S_{t},\{\beta_{t,i} \}^5_{i=1}, h_{t,R}, \{ h_{t,i}\}^5_{i=1})$ (blue curve).
    \end{tablenotes}
    \label{fig:tc_state_space_curves}
\end{figure}

\autoref{fig:tc_state_space_curves} illustrates the estimated value of the four penalty functions in proportion to that obtained with DH-L, $\rho(\xi_{T}^{\delta})/\rho(\xi_{T}^{\text{DH-L}})$, across different transaction cost rates for all agents. These numerical results reveal that RL agents outperform DH-L for all considered state space configurations, as the relative values of the penalty functions are substantially lower than the reference line value of 1.

Moreover, these results confirm that including the portfolio value in the full state space is unnecessary for our approach, as metrics under the reduced state space (green curve) show lower values than those achieved by agents using the full state space (red curve). This is consistent with the work of \cite{buehler2019deep}, \cite{cao2020deep}, \cite{buehler2022deep}, and \cite{cao2023gamma}, where RL techniques are applied in the hedging context without considering the portfolio value as a state variable.

In the absence of transaction costs, \autoref{fig:tc_state_space_curves} confirm that the inclusion of $\delta_{t}$ in the reduced state space is unnecessary, as the incremental performance with and without it is negligible. Conversely, in the presence of transaction costs, agents trained under the reduced state space $(\delta_{t},\tau_{t},S_{t},\{\beta_{t,i} \}^5_{i=1}, h_{t,R})$ demonstrate superior performance (green curve), and the outperformance with respect to benchmarks becomes more pronounced as the transaction cost rate increases. This observation underscores the importance of considering previous positions in the underlying when optimizing rebalancing actions, as the former is indicative of transaction costs to be paid for the various possible actions.

The superiority of RL agents trained on the reduced state space can be explained by the bias-variance dilemma, where the informational content provided by some of the variables (reduction in bias) is insufficient to compensate for additional complexity and instability (variance) they cause during training. 
This is seemingly why removing the IV parameters' variances $\{ h_{t,i}\}^5_{i=1}$ from the state space increases the performance in this experiment. Our numerical results indicate that the network architecture considered for the agents acting in our proposed high-dimensional environment does not require the full state space. In fact, the performance of the agents deteriorates for some transaction cost levels when the volatilities of the IV coefficients are included, as shown in Figure \ref{fig:tc_state_space_curves}, and the time required to converge to optimal solutions during training increases by an average of 190\% compared to RL agents under the reduced state space across all experiments. 

In all subsequent experiments, we consider the reduced state space $(\delta_{t},\tau_{t},S_{t},\{\beta_{t,i} \}^5_{i=1}, h_{t,R})$ in the presence of transaction costs and the state space $(\tau_{t},S_{t},\{\beta_{t,i} \}^5_{i=1}, h_{t,R})$ in the absence of transaction costs to enhance parsimony. 

\subsection{Benchmarking of hedging strategies}

In this section, we compare the performance of RL agents with classic hedging approaches. Aggregated results are presented in Section \ref{subsub:benchmarking_general}, while the segmentation among various states of the economy is shown in Section \ref{subsub:benchmarking_economy}. The following subsections break down the performance of the RL agents with respect to moneyness and maturity of the hedged option, or to transaction cost levels, demonstrating the robustness of the RL approach. Finally, Section \ref{subsub:benchmarking_realpaths} outlines a comparison of performance over historical paths spanning from January 5, 1996, to December 31, 2020.

\subsubsection{Benchmarking over randomized economic conditions}\label{subsub:benchmarking_general}

We begin by comparing the hedging performance of the benchmarks and RL agents trained under the four penalty functions: CVaR$_{95\%}$, CVaR$_{99\%}$, MSE, and SMSE. This comparison is performed in terms of estimated values of all penalty functions and the average Profit and Loss, $\mbox{Avg P\&L} = \mathbb{E}[ -\xi_{T}^{\delta}]$, across all paths in a test set. Additionally, we employ the CVaR deviation measure, defined as $\mbox{CVaR}_{\alpha}(\xi_{T}^{\delta} - E[\xi_{T}^{\delta}])$, as a deviation metric for the hedging error in the test set. In such test set, initial economic conditions (the initial value of state variables in the path) are sampled randomly among historical values from our sample. It therefore reflects aggregate performance across various economic conditions.
Our analysis focuses on hedging a European ATM call option with a maturity of $N=63$, under the assumption of no transaction costs, which is $\kappa=0$.

\begin{table}[h!]
    \centering
    \renewcommand{\arraystretch}{1.5}
    \caption{Aggregated hedging metrics for a short position in a 63-day ATM call option.}
    \label{table:aggregated_metrics}
    \begin{tabular}{{p{5.5cm}>{\centering\arraybackslash}p{1.3cm}>{\centering\arraybackslash}p{1.3cm}c>{\centering\arraybackslash}p{1.3cm}>{\centering\arraybackslash}p{1.3cm}>{\centering\arraybackslash}p{1.3cm}>{\centering\arraybackslash}p{1.3cm}}}
    \hline
    \multicolumn{1}{c}{} & \multicolumn{2}{c}{Benchmark} & & \multicolumn{4}{c}{Reinforcement Learning} \\
    \cline{2-3}\cline{5-8}
    Metric & DH & SI & & CVaR$_{95\%}$ & CVaR$_{99\%}$ & MSE & SMSE \\
    \hline
    Avg P\&L & 0.356 & {\bfseries 0.498} &   & 0.380 & 0.321 & 0.285 & 0.374 \\
    $\mbox{CVaR}_{\alpha}(\xi_{T}^{\delta} - E[\xi_{T}^{\delta}])$ & 2.159 & 3.005 &   & {\bfseries 1.646} & 1.720 & 1.766 &  1.652 \\
    CVaR$_{95\%}$ & 1.803 & 2.507 &   & {\bfseries 1.266} & 1.399 & 1.481 & 1.278 \\
    CVaR$_{99\%}$ & 3.442 & 4.351 &   & 2.312 & {\bfseries 2.198} & 2.625 & 2.245 \\
    MSE & 0.898 & 1.818 &   & 1.243 & 1.362 & {\bfseries 0.657} & 1.018 \\
    SMSE & 0.298 & 0.564 &   & 0.184 & 0.214 & 0.183 & {\bfseries 0.151} \\
    \hline
    Avg P\&L/$\mbox{CVaR}_{\alpha}(\xi_{T}^{\delta} - E[\xi_{T}^{\delta}])$ & 0.165 & 0.166 &   & {\bfseries 0.231} & 0.187 & 0.161 & 0.227 \\
    \hline
    \end{tabular}
    \begin{tablenotes}
    \item \footnotesize Results are computed using 100,000 out-of-sample paths in the absence of transaction costs ($\kappa = 0\%$). Agents are trained under the reduced state space $(\tau_{t},S_{t},\{\beta_{t,i} \}^5_{i=1}, h_{t,R})$ according to the conditions outlined in Section \ref{subsub:network_architecture}. 
    DH stands for delta hedging, whereas SI is the smile-implied delta hedging. 
    \end{tablenotes}
\end{table}

\autoref{table:aggregated_metrics} presents hedging performance metrics attained by both benchmarks and RL agents. 
Every RL agent achieves the lowest value for the corresponding metric which they used as objective function during training, which is expected.
Furthermore, the numerical results indicate that RL agents provide hedging strategies that are much less risky than benchmarks, as evidenced by metrics such as CVaR$_{95\%}$, CVaR$_{99\%}$, and SMSE. In particular, when computing the variation rate between the estimated value of each of these metrics obtained by RL agents relative to the value obtained by DH, we observe an average reduction rate across RL agents of 24\%, 31\%, and 38\%, respectively. Similarly, when comparing with SI delta strategies, we observe average reductions across RL agents of 45\%, 46\%, and 67\%, respectively. Moreover, in terms of the MSE metric, the MSE agent demonstrates significant superiority over benchmarks, reducing 26\% and 63\% compared to DH and SI delta strategies, respectively. 

Regarding average Profit and Loss (P\&L), SI delta yields higher profitability; however, it also entails increased risk. In contrast, RL agents trained using CVaR$_{95\%}$, CVaR$_{99\%}$, and SMSE as penalty functions obtain lower values compared to SI delta strategies, but they reach a more favorable trade-off between profitability and risk management as shown by the ratio \mbox{Avg P\&L/$\mbox{CVaR}_{\alpha}(\xi_{T}^{\delta} - E[\xi_{T}^{\delta}])$}, which yields greater values for these agents. Additionally, RL agents achieve lower risk than both the DH and SI delta hedging strategies, which highlights enhanced risk management. 

\subsubsection{Impact of the state of the economy on performance}\label{subsub:benchmarking_economy}

In this section, we analyze the performance metrics of hedging strategies across clusters of paths representing different economic states, as detailed in \autoref{tab:time_frames}. A path is assigned to a cluster if its state variable initial values are drawn from the subset of dates corresponding to that cluster. This approach helps isolate the impact of economic conditions on performance and assess the robustness of RL hedging strategies.
Again, we hedge an ATM European call option with a maturity of $N=63$ days under the assumption of no transaction cost.

\begin{table}[H]
    \centering
    \caption{Clusters of dates representing different time periods.}
    \begin{tabular}{>{\arraybackslash}m{6.5cm} >{\centering\arraybackslash}m{4.7cm}>{\centering\arraybackslash}m{4.7cm}}
        \toprule
        Period & Time frames & Avg option price \\
        \midrule
        Regular 1 & 05/01/1996 - 28/02/1997 & \$3.849 \\
        Crisis 1 (Asian financial crisis) & 03/03/1997 - 31/12/1998 & \$5.351 \\
        Regular 2 & 04/01/1999 - 31/12/1999 & \$5.301 \\
        Crisis 2 (Dot-com bubble crisis) & 03/01/2000 - 31/12/2002 & \$5.758 \\
        Regular 3 & 02/01/2003 - 31/12/2007 & \$3.877 \\
        Crisis 3 (Global financial crisis) & 02/01/2008 - 31/12/2009 & \$8.009 \\
        Regular 4 & 04/01/2010 - 31/01/2020 & \$3.728 \\
        Crisis 4 (Covid-19 pandemic crisis) & 03/02/2020 - 31/12/2020 & \$6.486 \\
        \bottomrule
    \end{tabular}
    \label{tab:time_frames}
    \begin{tablenotes}
    \item \footnotesize These periods aim to approximate different states of the economy to highlight the performance of our approach within time windows capturing the financial fluctuations characteristic of each economic crisis. The column "Avg option price" displays the average option price of an ATM call option  with maturity of 63 days per cluster.
    \end{tablenotes}
\end{table}

\begin{figure}[H]\centering
\caption{Hedging metrics for a short position in an ATM call option with maturity of 63 days under different states of the economy.}
\label{fig:states_economy}
\includegraphics[width=17cm]{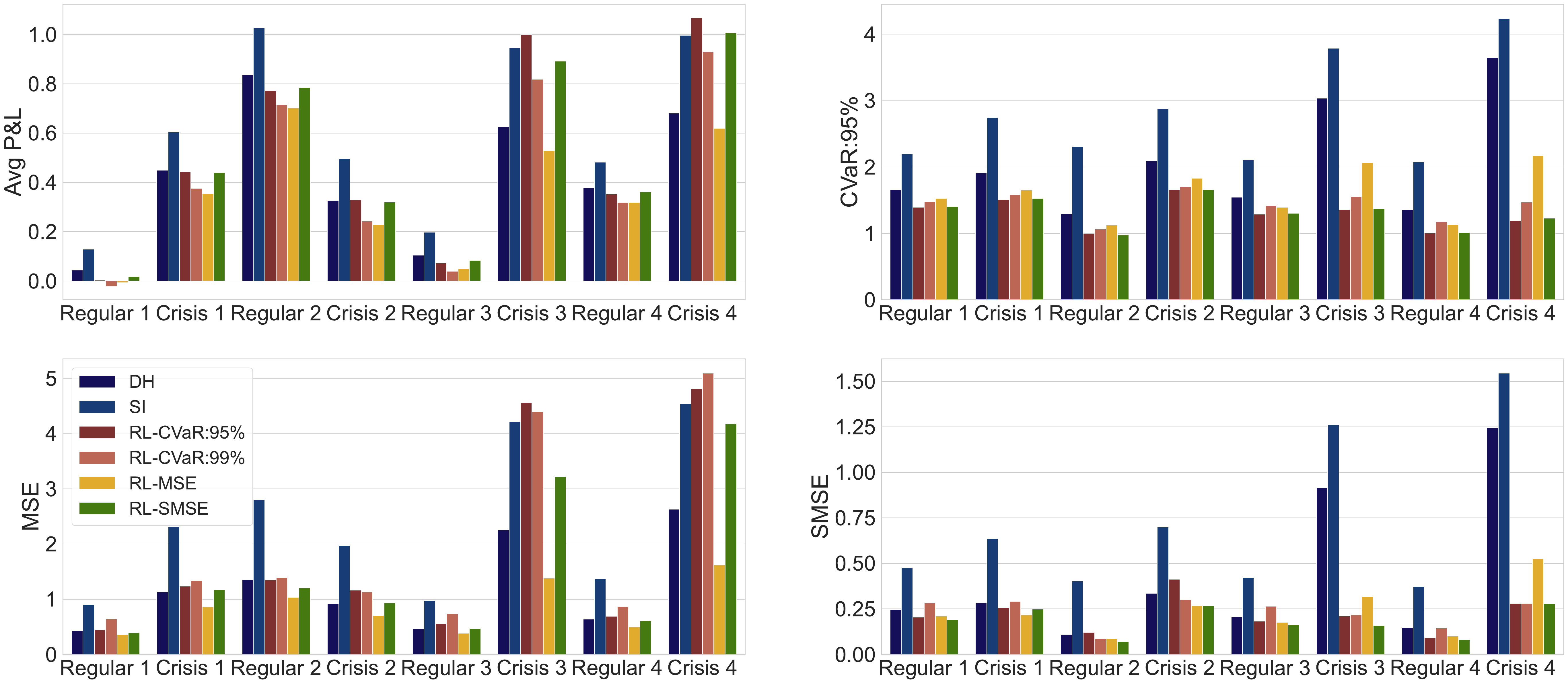}
\begin{tablenotes}
\item \footnotesize Results are computed using 100,000 out-of-sample paths in the absence of transaction costs ($\kappa = 0\%$). Agents are trained under the reduced state space $(\tau_{t},S_{t},\{\beta_{t,i} \}^5_{i=1}, h_{t,R})$ according to the conditions outlined in Section \ref{subsub:network_architecture}. 
The results are organized chronologically across the periods outlined in \autoref{tab:time_frames}.
\end{tablenotes}
\end{figure}

Our numerical results, illustrated in \autoref{fig:states_economy}, indicate that both benchmarks and RL agents are sensitive to the environment, showcasing better performance during regular periods compared to financial crises in terms of risk, as shown by the two right plots showing the CVaR$_{95\%}$ and SMSE metrics. Additionally, both approaches tend to offer more profitable strategies during crisis periods and periods of high volatility (for instance, even though Regular period 2 is not labeled as a crisis period, it is characterized by economic recovery with significant market fluctuations). This trend can be explained by the fact that higher volatility during these periods leads to a higher risk premium, increasing the initial portfolio value and thus the final P\&L, as highlighted by the Avg P\&L metric (see the top left panel of \autoref{fig:states_economy}). Regarding the MSE metric, the results align with the Avg P\&L metric, with both approaches showing superior performance during regular periods (except for Regular period 2 which exhibited high volatility), as higher profits are detrimental for the MSE metric penalizing upside risk.

The results that are segregated by time period align with the aggregated findings. Indeed, for any of the CVaR$_{95\%}$, MSE, and SMSE risk metrics, the RL agent which was trained with such risk metric as its objective function consistently exhibits lower values than benchmarks across all periods. These differences are particularly notable during crisis periods, where DH and SI delta tend to be riskier, while RL agents demonstrate greater stability. The higher values in the MSE statistic of agents trained under CVaR and SMSE can be attributed to the fact that these objective functions only penalize hedging losses, allowing agents to seek positive returns on average. This is consistent with the Avg P\&L panel, where CVaR and SMSE agents tend to display more profitable strategies than the MSE agent.

These results demonstrate the robustness of the RL approach across various environments, exhibiting greater stability in scenarios with extreme behavior and improving hedging performance in terms of the penalty functions considered in the hedging problem. Additionally, they support our initial findings, indicating that RL agents do not significantly sacrifice profitability even during extreme market conditions, a phenomenon that could be attributed to higher initial option prices during these periods, as shown in \autoref{tab:time_frames}.

\subsubsection{Impact of moneyness level on performance}\label{subsub:benchmarking_moneyness}

\begin{figure}[H]\centering
\caption{Hedging metrics for a short position in OTM, ATM and ITM call options with a maturity of 63 days.}
\includegraphics[width=17cm]{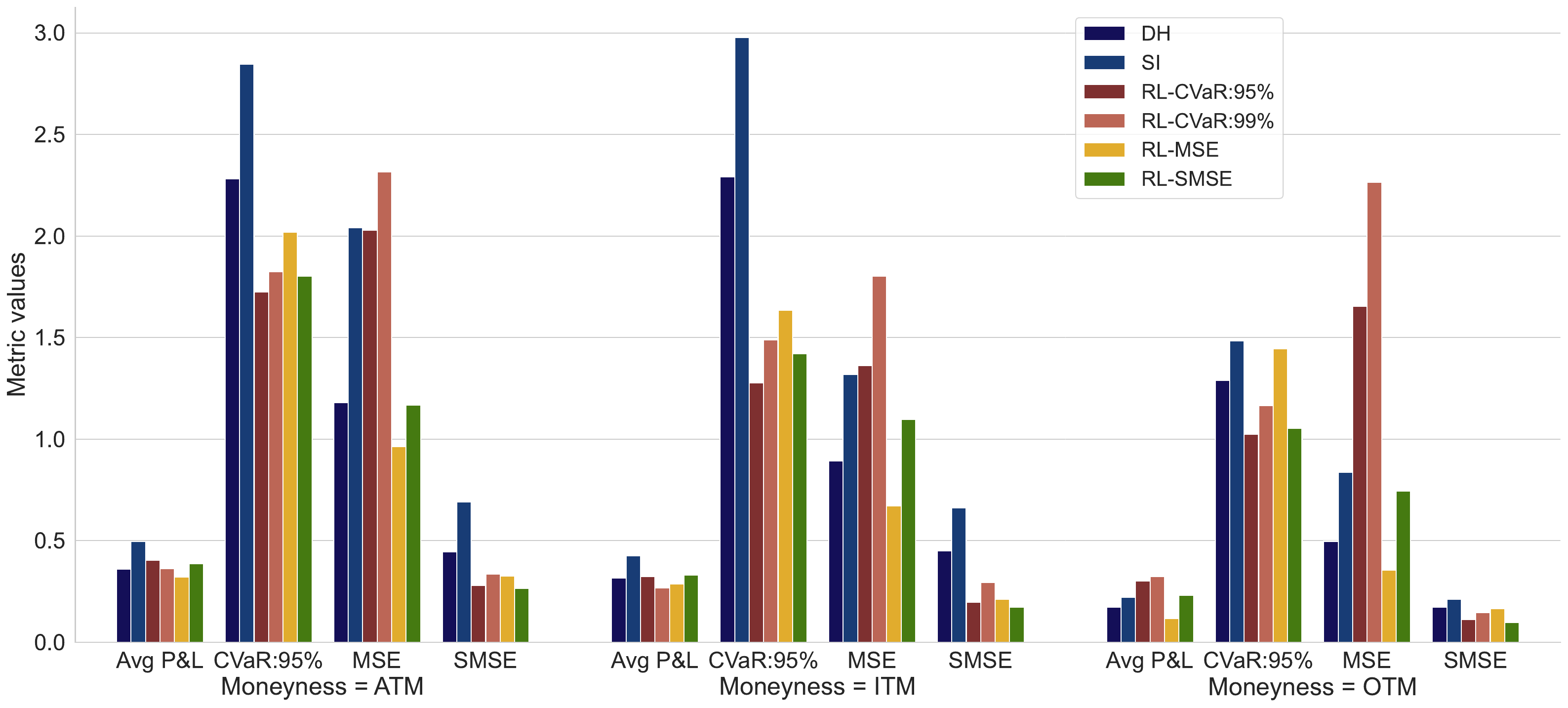}
\begin{tablenotes}
\item \footnotesize Results are computed using 100,000 out-of-sample paths in the absence of transaction costs ($\kappa = 0\%$). Agents are trained under the reduced state space $(\tau_{t},S_{t},\{\beta_{t,i} \}^5_{i=1}, h_{t,R})$ according to the conditions outlined in Section \ref{subsub:network_architecture}. 
The average option price stands at \$0.59 for OTM options, \$3.89 for ATM options, and \$11.37 for ITM options.
\end{tablenotes}
\label{fig:moneyness}
\end{figure}

We investigate the robustness of our approach with respect to the moneyness level of the option being hedged by including OTM and ITM call options with a maturity of $N=63$ days in our analysis. Our findings, depicted in \autoref{fig:moneyness}, demonstrate that RL consistently outperforms benchmarks with respect to the objective function considered during the training process, regardless of the option moneyness.

In line with our previous experiments, we observe that SI delta tends to offer more profitable strategies compared to benchmarks, while DH demonstrates a similar level of profitability for ATM and ITM options. However, this trend does not hold for OTM options, where RL agents not only exhibit better risk management but also yield more profitable strategies on average (refer to CVaR and SMSE agents for OTM options in \autoref{fig:moneyness}). The discrepancy can be attributed to the fact that RL agents trained under CVaR and SMSE do not track the option value nor penalize gains at maturity, allowing agents to profit from OTM paths at maturity. In contrast, DH and SI are option tracking methodologies that aim to replicate the option value at maturity, thereby reducing potential gains for OTM options at maturity. The latter is consistent with the RL agent trained under the MSE, which achieves the minimum MSE value for OTM options and the lowest Avg P\&L, as it penalizes both gains and losses.

\subsubsection{Impact of option maturity on performance}\label{subsub:benchmarking_maturity}

\begin{figure}[h]\centering
\caption{Hedging metrics for a short position in ATM call options with maturities of 21, 63 and 126 days.}
\includegraphics[width=17cm]{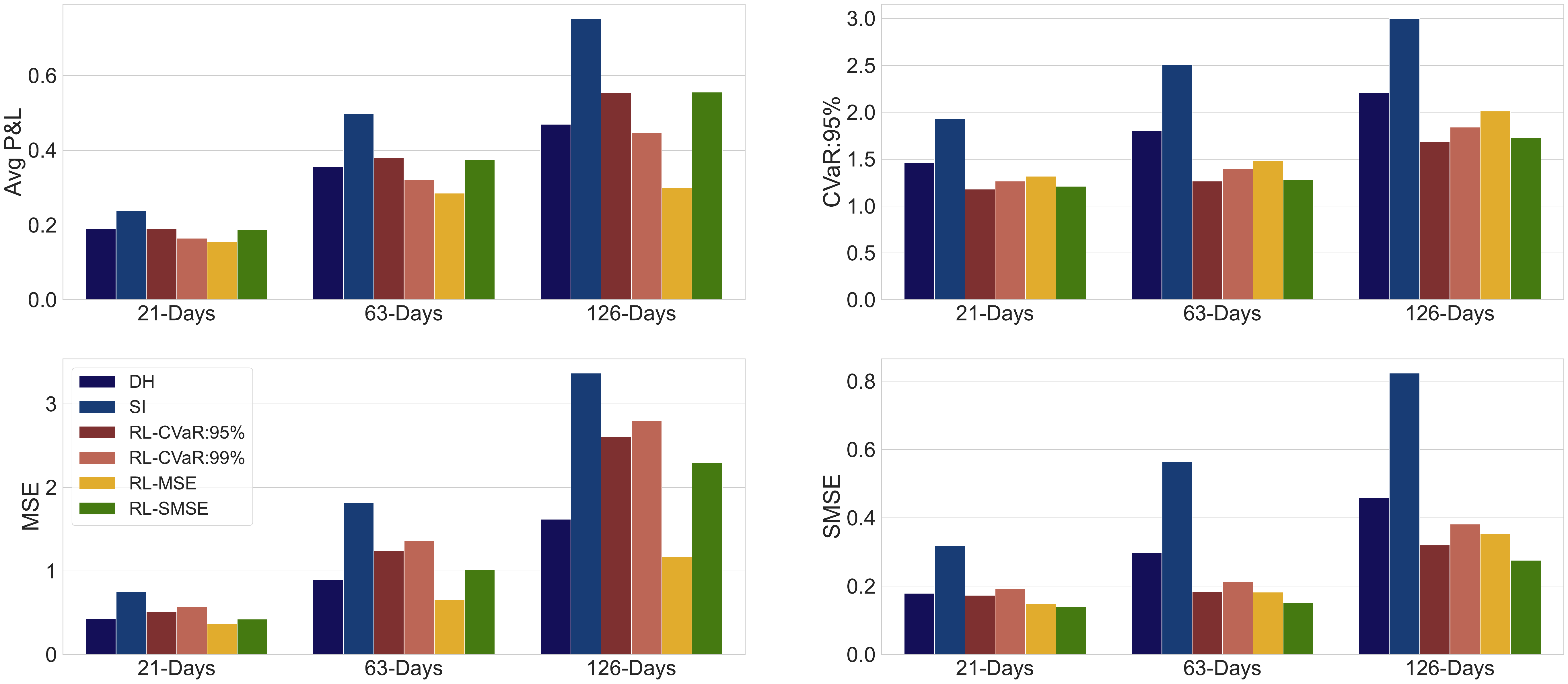}
\begin{tablenotes}
\item \footnotesize Results are computed using 100,000 out-of-sample paths in the absence of transaction costs ($\kappa = 0\%$). Agents are trained under the reduced state space $(\tau_{t},S_{t},\{\beta_{t,i} \}^5_{i=1}, h_{t,R})$ according to the conditions outlined in Section \ref{subsub:network_architecture}. Average prices are \$2.15, \$3.89 and \$5.65 for options with maturities of 21 days, 63 days and 126 days, respectively.
\end{tablenotes}
\label{fig:maturity}
\end{figure}

As a third test to assess the robustness of our approach, we compare the performance of RL agents against benchmarks when hedging ATM call options with different maturities: 21, 63, and 126 days, in absence of transaction costs. Our results, illustrated in \autoref{fig:maturity}, show that the average profitability of hedging strategies increases with option maturity (see Avg P\&L in the top left panel). However, the risk also increases with maturity for all hedging strategies (refer to CVaR$_{95\%}$ and SMSE depicted in the panels positioned at the top and bottom right corners), as expected.

These results also demonstrate consistent behavior across all maturities for all hedging metrics, displaying a uniform distribution among the different strategies across all maturity levels. In line with our previous experiments, RL agents consistently outperform benchmarks with respect to asymmetric penalty functions for all maturities, regardless of the penalty function used during training. Furthermore, in the case of the MSE metric (bottom left panel in \autoref{fig:maturity}), the RL agent trained under that penalty function displays the lowest value across all maturities.

\subsubsection{Impact of transaction costs on performance}\label{subsub:benchmarking_tc}

We now investigate scenarios where hedgers encounter transaction costs, which is meant to reflect more realistic hedging scenarios. Specifically, we analyze the hedging effectiveness of RL agents in comparison to benchmarks, including the DH-L strategy. Our examination focuses on hedging an ATM European call option with a maturity of $N=63$ days across varying levels of transaction costs assumed to be 0.05\%, 0.5\% and 1\%.

\begin{figure}[h]\centering
    \caption{Hedging metrics for a short position in an ATM call option with a maturity of 63 days under different transaction cost levels.}
    \includegraphics[width=17cm]{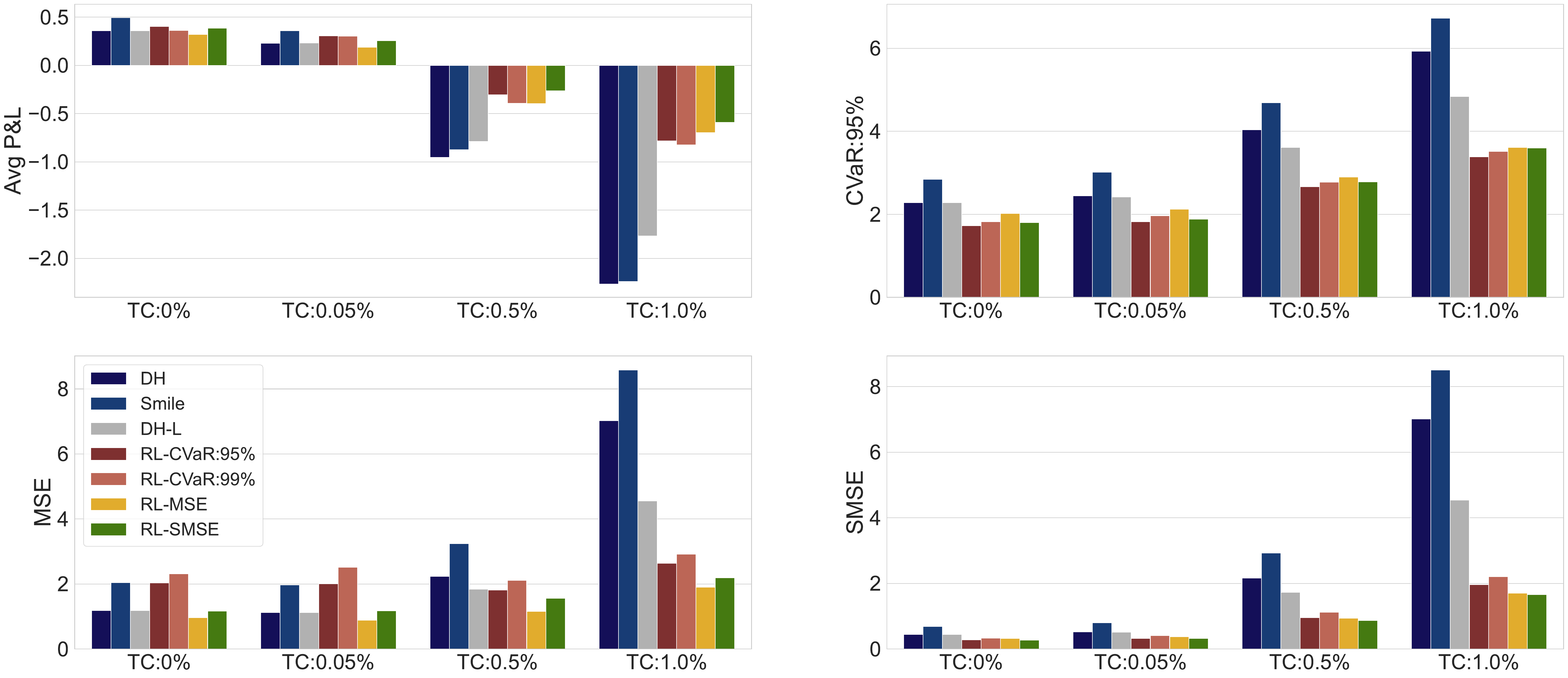}
    \begin{tablenotes}
    \item \footnotesize Results are computed using 100,000 out-of-sample paths under the reduced state space $(\tau_{t},S_{t},\{\beta_{t,i} \}^5_{i=1}, h_{t,R})$ without transaction costs, and under $(\delta_{t},\tau_{t},S_{t},\{\beta_{t,i} \}^5_{i=1}, h_{t,R})$ in the presence of transaction cost. Agents are trained according to the conditions outlined in Section \ref{subsub:network_architecture}. The average option price is \$3.89 with a standard deviation of \$1.29. The acronym TC stands for transaction cost. DH Black-Scholes delta hedging, DH-L for Leland delta hedging and SI for smile-implied delta hedging.
    \end{tablenotes}
    \label{fig:transaction_costs}
\end{figure}

\autoref{fig:transaction_costs} illustrates the performance of benchmarks and RL agents, with the four panels depicting the Avg P\&L, CVaR$_{95\%}$, MSE, and SMSE metrics. In general, as expected, the performance of each strategy tends to deteriorate as transaction costs increase. However, performance drops are more pronounced for benchmarks than for RL agents. As anticipated, RL agents consistently outperform benchmarks in terms of downside risk (observe CVaR$_{95\%}$ and SMSE metrics depicted in the panels positioned at the top and bottom right corners of \autoref{fig:transaction_costs}) across all transaction cost levels, regardless of the penalty function used during training. Furthermore, while DH-L adjusts the strategies in terms of the transaction cost level and performs better than the other benchmarks, RL strategies display superior metrics, highlighting their better capacity to adapt to different transaction cost levels.

In contrast to our previous results where SI delta provides more profitability, the results regarding the Avg P\&L metric (see top left panel in \autoref{fig:transaction_costs}) indicate that this profitability is influenced by the inclusion of transaction costs, making the SI delta hedging less profitable strategy as transaction costs increase, whereas RL agents tend to exhibit much lower average losses due to their adaptability. In terms of MSE, benchmarks exhibit a more sensitive behavior than the RL agents regarding the increase in transaction costs, especially when the transaction costs are set at 0.5\% and 1\% (see the bottom-left panel of \autoref{fig:transaction_costs}), irrespective of the penalty function considered during training.

\autoref{tab:transaction_cost} displays descriptive statistics about hedging costs, defined as the sum of discounted transaction costs over a path,
\begin{equation}
    \sum_{t=0}^{T-1}e^{-r\Delta t}\kappa S_{t}\mid \delta_{t+1} - \delta_{t} \mid,
\end{equation}
across different transaction cost levels. In the cases of DH and SI delta strategies, which overlook transaction costs, we observe a higher cost of hedging due to significant variations in the position of the underlying asset during rebalancing across different time steps, thereby leading to lower profitability as transaction costs increase. This trend is evident in the case of SI delta which incurs the highest average hedging cost, followed by DH and DH-L, when the transaction cost rate is set to 0.5\% and 1\%. These results consistently align with the performance differences observed in \autoref{fig:transaction_costs},  where benchmarks are associated with the highest losses.

\begin{table}[h]
\centering
\renewcommand{\arraystretch}{1.5}
\caption{Hedging costs when hedging a short ATM call option position with a maturity of $N=63$ days under various transaction cost levels.}
\begin{tabular}{{p{1.7cm}>{}p{1.7cm}>{\centering\arraybackslash}p{1.3cm}>{\centering\arraybackslash}p{1.3cm}>{\centering\arraybackslash}p{1.3cm}c>{\centering\arraybackslash}p{1.3cm}>{\centering\arraybackslash}p{1.3cm}>{\centering\arraybackslash}p{1.3cm}>{\centering\arraybackslash}p{1.3cm}}}

\hline
 \multicolumn{2}{c}{} & \multicolumn{3}{c}{Benchmark - Delta} & & \multicolumn{4}{c}{Reinforcement Learning} \\

\cline{3-5}\cline{7-10}

$\kappa$ & Metric & DH & SI & DH-L & & CVaR$_{95\%}$ & CVaR$_{99\%}$ &MSE & SMSE  \\
\hline
0.05\% & Mean &0.138 & 0.142 & 0.136 &   & 0.112 & {\bfseries 0.108} & 0.150 & 0.113 \\
 & Std & 0.041 & 0.048 & 0.040 &   & 0.024 & {\bfseries 0.021} & 0.042 & 0.025 \\
0.5\% & Mean & 1.376 & 1.420 & 1.223 &   & 0.751 & 0.788 & 0.767 & {\bfseries 0.674} \\
 & Std & 0.412 & 0.482 & 0.317 &   & 0.137 & 0.137 & 0.205 & {\bfseries 0.127} \\
1\% & Mean & 2.753 & 2.839 & 2.260 &   & 1.245 & 1.259 & 1.139 & {\bfseries 1.031} \\
 & Std & 0.824 & 0.964 & 0.535 &   & 0.203 & 0.212 & 0.254 & {\bfseries 0.188} \\
\hline
\end{tabular}
\label{tab:transaction_cost}
\begin{tablenotes}
\item \footnotesize The average cost of hedging is computed by evaluating the hedging cost across 100,000 out-of-sample independent paths. Transaction cost levels are assumed to be proportional to the trade size. The lowest values across all hedging strategies are highlighted in bold. The average option price is \$3.89 with a standard deviation of \$1.29. Std stands for standard deviation. 
\end{tablenotes}
\end{table}

Furthermore, although none of the agents directly aim to minimize transaction costs, we observe that RL agents with asymmetric penalty functions tend to offer hedging strategies with lower costs, as shown by the highlighted values in \autoref{tab:transaction_cost}. This underscores the robustness of RL agents in addressing transaction costs, as they minimize the penalty function and thereby indirectly reduce the cost of hedging. Conversely, the RL agent trained under the MSE criterion tends to exhibit higher average costs compared to the other agents. This can be attributed to the MSE agent penalizing both negative and positive returns, creating additional turnover in some situations to avoid large profits. 

\subsection{Global importance of IV surface factors}\label{sub:global_inportance}

We now investigate to what extent the risk factors characterizing the IV surface described in Equation \eqref{5factormodel_linear} contribute to total performance of our model. 
We employ the Shapley Additive Global Importance (SAGE) methodology, introduced by \cite{covert2020understanding}, to evaluate their impact. 
Specifically, let $\rho\left(\xi_{T}^{\tilde{\delta}_{\theta}(\cal{X})}\right)$ be the risk measure when the model is trained with the state space $\cal{X}$. The amount of risk reduction achieved by adding the state variables $(\{\beta_{t,i}\}^5_{i=1}, h_{t,R})$ to a baseline model \(\tilde{\delta}_{\theta}\) with state variables $(\tau_{t}, S_{t})$ is
\begin{align*}
        \rho\left(\xi_{T}^{\tilde{\delta}_{\theta}(\tau_{t}, S_{t})}\right) 
        &- \rho\left(\xi_{T}^{\tilde{\delta}_{\theta}(\tau_{t}, S_{t}, \{\beta_{t,i}\}^5_{i=1}, h_{t,R})}\right)= \sum_{j \in \{\{\beta_{t,i}\}^5_{i=1}, h_{t,R} \}} \mathcal{C}_{j}\quad 
    \end{align*}
where $\mathcal{C}_{j}$, the contribution of variable $j$ to the total risk reduction, is
\begin{equation*}
    \mathcal{C}_{j}= \sum_{\mathcal{X} \subseteq \{ \tau_{t}, S_{t}, \{\beta_{t,i}\}^5_{i=1}, h_{t,R}\} \setminus \{j\}} \frac{|\cal{X}|! (5 - |\cal{X}|)!}{6!} \left[ \rho\left(\xi_{T}^{\tilde{\delta}_{\theta}(\cal{X})}\right) - \rho\left(\xi_{T}^{\tilde{\delta}_{\theta}(\cal{X}, j)}\right) \right].
\end{equation*}
Exact contributions $\mathcal{C}_{j}$ cannot be negative, although their estimates sometimes are. Negative values may occur for the contributions we present that are evaluated out-of-sample.
Our numerical experiments involve training RL agents using four penalty functions: CVaR\(_{95\%}\), CVaR\(_{99\%}\), MSE, and SMSE. We analyze the global importance of these state variables across different moneyness and maturities to comprehend their impact on the performance of the RL agents. The global importance of state variables is normalized by the risk reduction achieved by the respective RL agent to present contributions in the same order of magnitude: the relative global importance is

\begin{equation}\label{eq:relative_shap}
    \frac{\mathcal{C}_{j}}{\rho\left(\xi_{T}^{\tilde{\delta}_{\theta}(\tau_{t}, S_{t})}\right) 
    - \rho\left(\xi_{T}^{\tilde{\delta}_{\theta}(\tau_{t}, S_{t}, \{\beta_{t,i}\}^5_{i=1}, h_{t,R})}\right)},\quad \mbox{for } j \in \{h_{t,R}\}\cup \{\beta_{t,i}\}^5_{i=1}.
\end{equation}

The relative global importance of the IV characteristic $\beta_{t,i}$ and the return conditional variance $h_{t,R}$ to the risk reduction depends on the moneyness and the time-to-maturity of the option to be hedged, as well as on the choice of risk measure.

\begin{figure}[h]\centering
\caption{Normalized global importance of state variables when hedging a European call options with a maturity $N=63$ days, across various moneyness levels.}
\includegraphics[width=16cm]{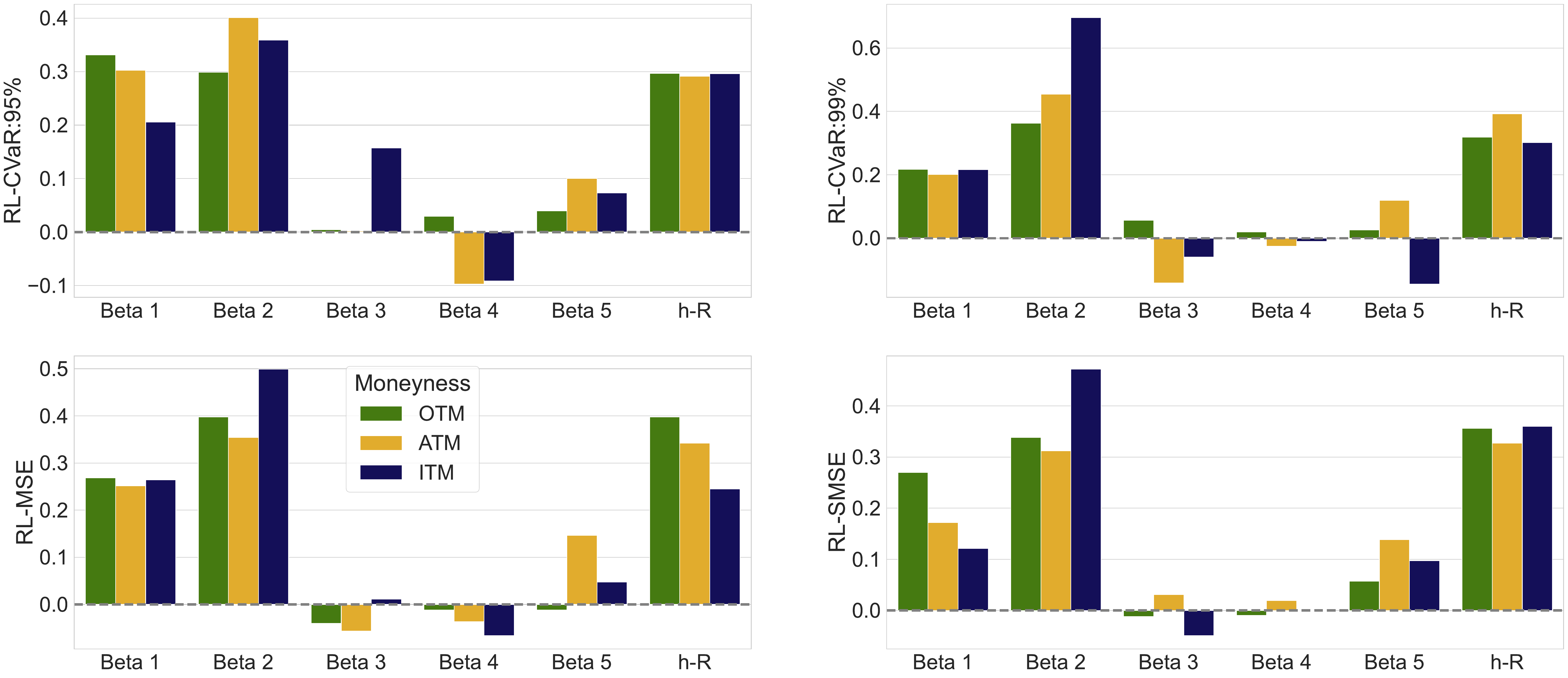}
\begin{tablenotes}
\item \footnotesize Results are computed using 100,000 out-of-sample paths in the absence of transaction costs ($\kappa = 0\%$). Each panel illustrates the Shapley values for all state variables $(\{\beta_{t,i}\}^5_{i=1}, h_{t,R})$ and different moneyness: OTM, ATM, and ITM. These results are shown for the four RL agents:  RL-CVaR$_{95\%}$, RL-CVaR$_{99\%}$, RL-MSE, and RL-SMSE. The Shapley values are normalized by the risk reduction achieved by the respective agent according to Equation \eqref{eq:relative_shap}.
\end{tablenotes}
\label{fig:sage_moneyness}
\end{figure}

\autoref{fig:sage_moneyness} studies the case of 63-day-to-maturity call options. It illustrates the relative contribution of each state variable across moneyness levels. Overall, the conditional variance of the underlying asset returns, the long-term ATM level $\beta_1$ and the time-to-maturity slope $\beta_2$ of the IV surface play a major role, no matter what risk measure or moneyness is considered. This underscores that RL agents utilize both the historical variance process and market expectations of future volatility to adjust positions in the underlying asset. The moneyness slope, the smile attenuation and the smirk have a second order effect. 

\begin{figure}[h!]\centering
    \caption{Global importance of state variables when hedging a European ATM call options, across various maturities.}
    \includegraphics[width=16cm]{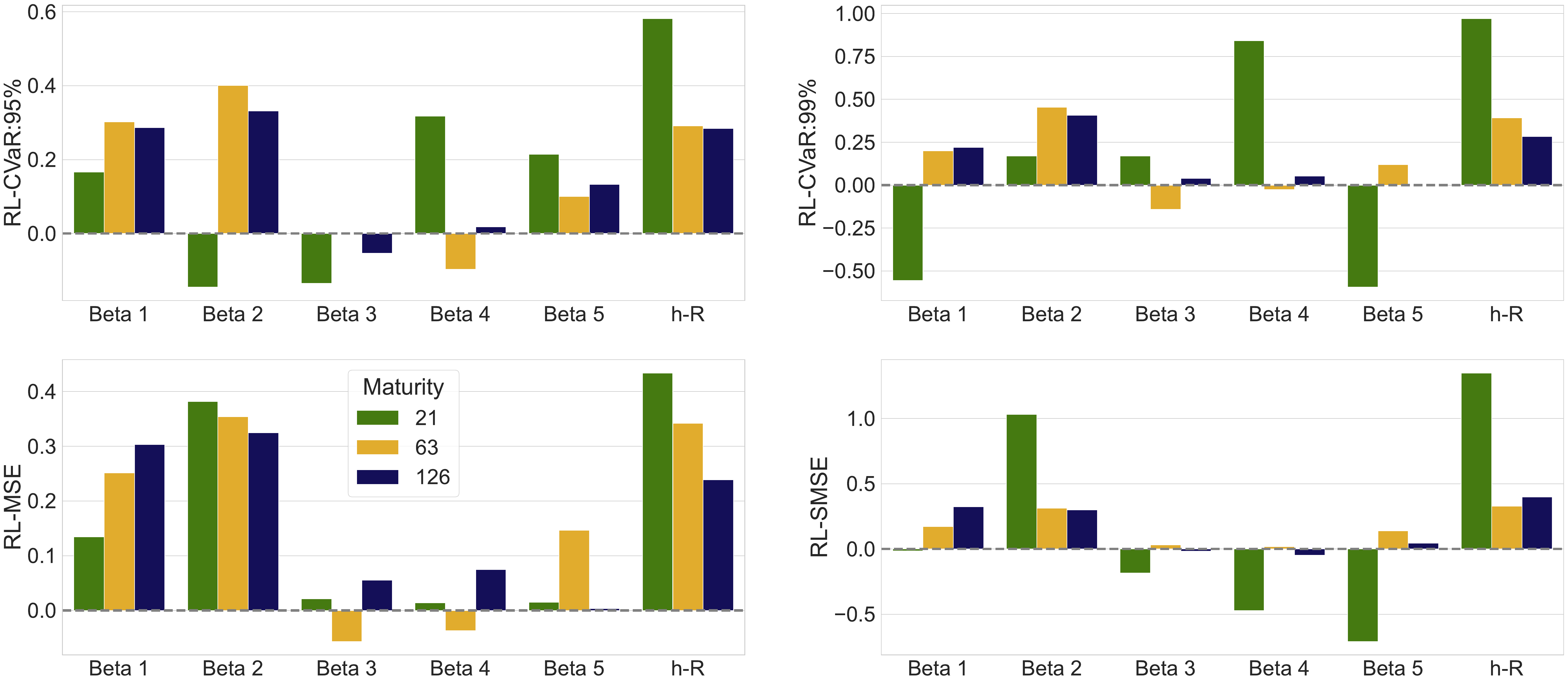}
    \begin{tablenotes}
    \item \footnotesize Results are computed using 100,000 out-of-sample paths in the absence of transaction costs ($\kappa = 0\%$). Each panel illustrates the Shapley values for all state variables under different maturities: 21, 63, and 126 business days. These results are shown for the four RL agents: CVaR$_{95\%}$, CVaR$_{99\%}$, MSE, and SMSE. The Shapley values are normalized by the risk reduction achieved by the respective agent according to Equation \eqref{eq:relative_shap}.
    \end{tablenotes}
    \label{fig:sage_maturity}
\end{figure}

\autoref{fig:sage_maturity} presents contributions to hedging risk reduction for three option maturities, namely 21, 63 or 126 business days. It underscores the persistent significance of the conditional variance process \( \{h_{t,R}\} \) across the various option maturities. Its contribution is even more important for short-term maturities. This reflects the fact that \( h_{t,R} \) has a direct impact on immediate market shocks.
Both $\beta_4$ and $\beta_5$ are additional contributors mostly for short-term options when CVaR risk measures are considered. Intuitively, high values of the smile attenuation ($\beta_4$) and the smirk ($\beta_5$) indicate a steep slope and a strong smirk for the short term smile which, in turn, induces a high exposure to tail risk for short term options.

\section{Backtesting} \label{subsub:benchmarking_realpaths}

We conduct out-of-sample backtesting experiments to assess the model’s performance on actual data. Specifically, we evaluate the effectiveness of our approach using traded option prices observed between December 31, 2020, and October 31, 2023, ensuring that the RL agents, trained on simulated data reflecting estimated dynamics from January 1996 to December 2020, are tested on realized market data they have never previously encountered.

All options enter the sample as soon as their remaining time-to-maturity is 63 business days, at which point a short position is taken and hedged.\footnote{
    To ensure liquidity and reliability throughout the hedging period, we retain only options that report non-zero values for best bid, trading volume, and open interest at the onset. These filters are based on the fields \texttt{best\_bid}, \texttt{volume}, and \texttt{open\_interest} from OptionMetrics. We include all options that meet the criteria within the test set period.}
    All retained options are European calls that are around-the-money, defined as having an underlying price within $\pm10\%$ of the strike.\footnote{Because option prices are proportional to the initial value of the underlying asset in our framework, we normalize all prices such that the initial value of the underlying is \$100. This allows us to replicate conditions similar to those used to train the RL algorithms, without any loss of generality.}
    This leaves us with 4,134 hedged contracts.
The state variables are reconstructed from historical data using \cite{Frnacois2022} and \cite{Frnacois2023}. In the presence of transaction costs, the state space is augmented to include the current portfolio position, resulting in $(\tau_t, S_t, \{\beta_{t,i}\}_{i=1}^5, h_{t,R}, \delta_t)$, where $\delta_t$ denotes the number of shares held at time $t$.


We compare the performance of benchmark strategies—DH, SI, and DH-L (with transaction costs)—against our RL-based approaches. We also include a benchmark RL agent that excludes the IV surface from its input features, namely $\{ \beta_{t,i}\}^5_{i=1}$. This restricted RL strategy exhibits weaker risk reduction in the simulation environment, as highlighted by the Shapley value decomposition from \cref{sub:global_inportance} which reveals a positive contribution from the incorporation of IV coefficients. We evaluate whether this performance gap persists in the out-of-sample backtest. All strategies are assessed using three risk metrics: MSE, SMSE, and CVaR$_{95\%}$.


\begin{figure}[H]\centering
    \caption{Backtest risk metrics on hedging errors, with and without transaction costs.}
    \includegraphics[width=17.5cm]{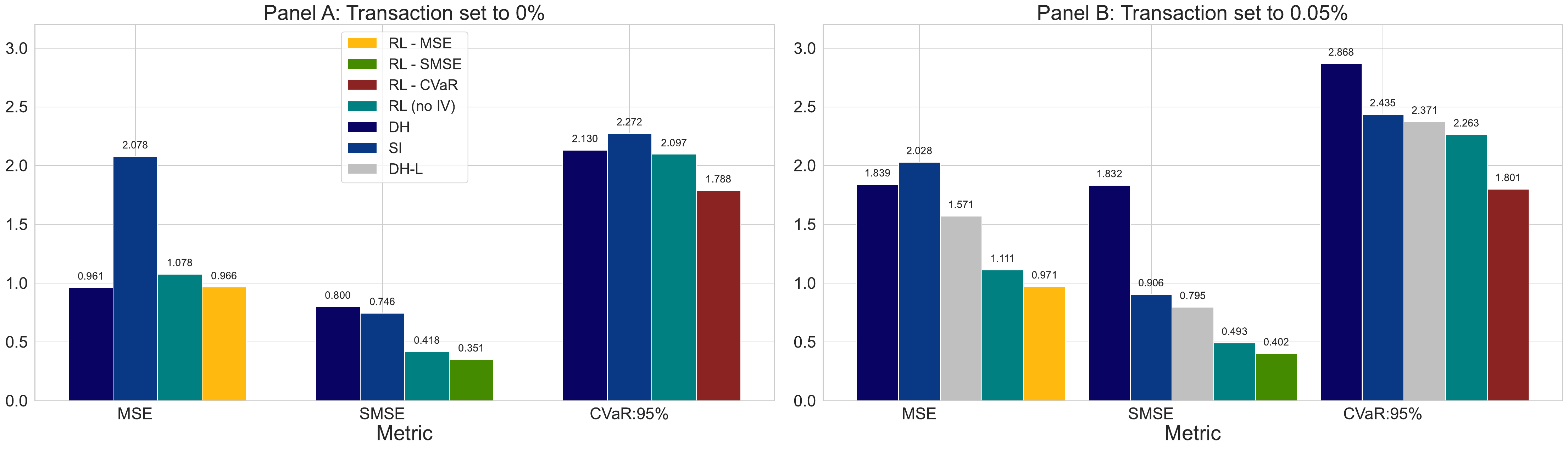}
    \begin{tablenotes}
    \item 
    \footnotesize Backtests are conducted on 4,134 around-the-money call options, using actual market prices observed between December 31, 2020 and October 31, 2023. 
    \end{tablenotes}
    \label{fig:backtest_metrics}
\end{figure}

As shown in \autoref{fig:backtest_metrics} the RL algorithm with IV information and the smile-implied delta hedging, both of which leverage the IV surface, deliver comparable performance in terms of MSE. They are closely followed by the RL algorithm with restricted information, while the practitioners' delta hedging lags significantly behind. When transaction costs are introduced, both RL algorithms outperform all variants of delta hedging strategies. When focusing on tail risk measures—SMSE and CVaR$_{95\%}$—RL strategies consistently outperform traditional benchmarks. Moreover, incorporating the IV surface information yields additional performance gains.



\begin{figure}[h!]
    \centering
    \caption{Backtest hedging error distributions, without transaction costs.}
    \includegraphics[width=17.5cm]{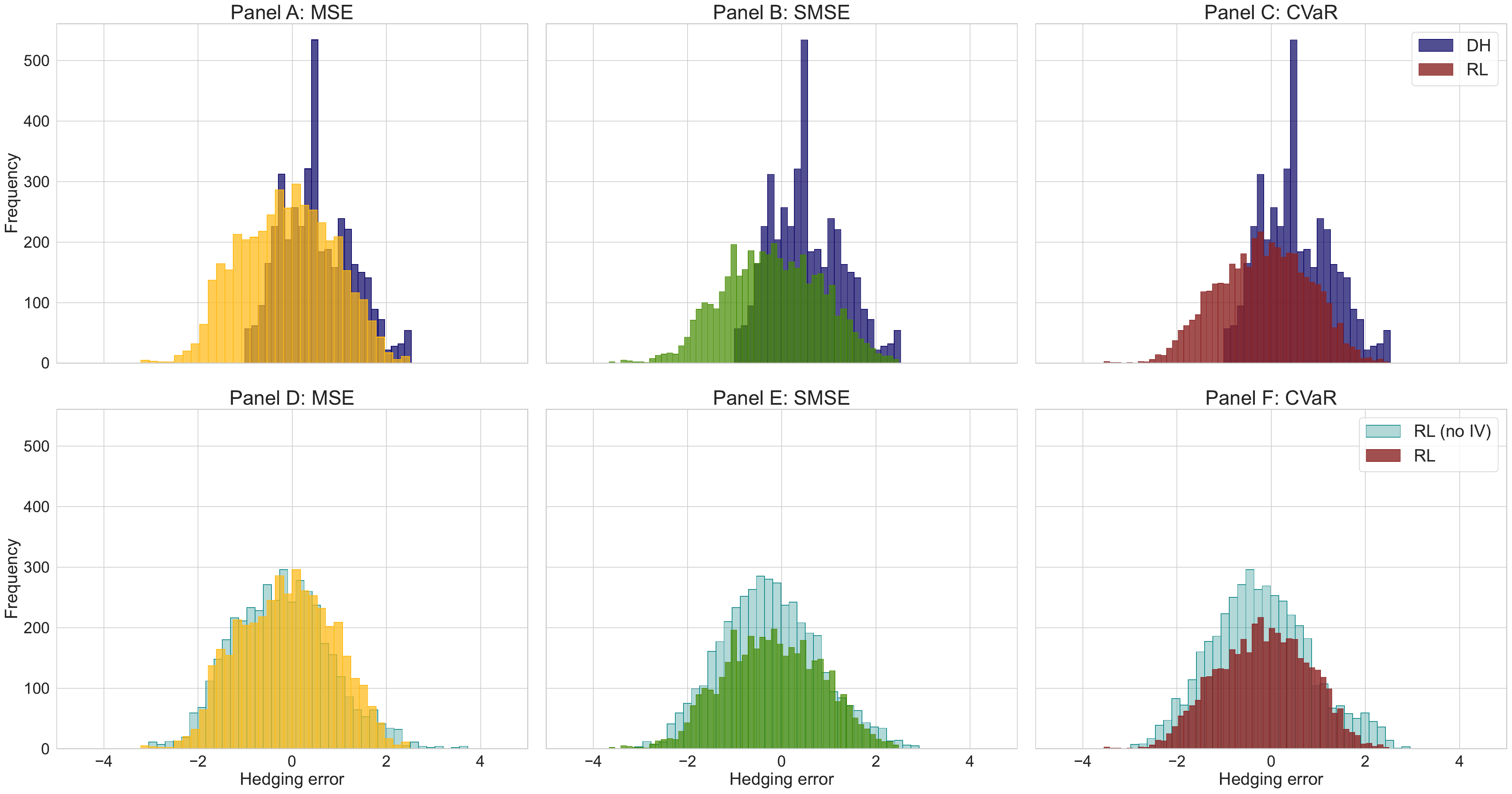}
    \begin{tablenotes}
    \item 
    \footnotesize The results are based on hedging 4134 around-the-money call options using real market prices observed between December 31, 2020 and October 31, 2023. No transaction costs are applied.
    \end{tablenotes}
    \label{fig:backtest_distribution}
\end{figure}

The superior performance of RL strategies over delta-based benchmarks is further illustrated in \autoref{fig:backtest_distribution}. When the risk metric is the MSE, the goal is to center the hedging error around zero. As shown in Panel A, the RL strategy is more successful in achieving such objective than the practitioner's delta hedge.

For asymmetric risk metrics such as SMSE and CVaR, the objective shifts toward limiting exposure to extreme losses. Panels B and C show that the RL strategies produce notably thinner right tails in the hedging error distributions, reflecting a more effective control of downside risk. A similar pattern is observed in panels D, E, and F, where RL strategies with the full market information also outperform their counterparts that exclude the IV surface information.

These findings suggest that RL-based hedging can serve as an effective risk management tool. However, the inclusion of relevant information from the implied volatility surface is required to unleash its full potential; its performance depends on the quality and completeness of the input data.


\section{Conclusion}\label{se:conclusion}

This study introduces a novel deep hedging framework that integrates forward-looking volatility information through a functional representation of the IV surface, combined with backward-looking conventional state variables. Our implementation employs deep policy gradient methods and utilizes a neural network architecture consisting of LSTM cells and FFNN layers to enhance training efficiency. Additionally, the architecture incorporates a budget constraint mitigating the incentive to gamble and enabling the agent to learn hedging strategies that are more effective for risk management.

Our approach consistently outperforms traditional benchmarks both in the absence and presence of transaction costs. The stability of our approach is assessed across various economic states using simulated data, demonstrating greater robustness than benchmarks in extreme scenarios such as economic crises.

The global importance analysis of IV factors confirms the significant enhancement of hedging performance in terms of risk reduction relative to the penalty functions. Our analysis underscores the critical importance of key factors such as the underlying asset return conditional variance process ($h_{R}$), the long-term ATM IV level ($\beta_{1}$) and the time-to-maturity slope ($\beta_{2}$). RL agents utilize both the historical variance process and market expectations of future volatility to adjust positions in the underlying asset. 
Out-of-sample backtests underscore the critical role of IV surface information in unlocking the full potential of RL-based hedging strategies.


{

\bibliographystyle{apalike}
\bibliography{references}  
}


\appendix

\section{Benchmarks}\label{appen:benchmarks}

The three benchmarks outlined in this appendix operate under the premise that the implied volatilities follow the IV model introduced in Equation \eqref{5factormodel_linear}.


The \cite{BlackScholes} pricing formula for an European call option is:
$$\mbox{Call}_{t}=S_{t} \mbox{e}^{-q_{t}\tau_{t}}\Phi(d_{t})-K e^{ -r_{t}\tau_{t}}\Phi(d_{t} - \sigma (M_{t},\tau_{t},\beta_{t})\sqrt{\tau_{t}})$$

where $d_{t}=\frac{\mbox{log}\left( \frac{S_{t}}{K} \right)+\left( r_{t}-q_{t}+\frac{1}{2}\sigma (M_{t},\tau_{t},\beta_{t})^2 \right)\tau_{t}}{\sigma (M_{t},\tau_{t},\beta_{t}) \sqrt{\tau_{t}}}$, $\sigma_t$ is the implied volatility of the option and $\Phi$ is the cumulative distribution function of the standard normal distribution. Moreover, the Black-Scholes delta is
\begin{equation*}
    \Delta_{t}^{BS} = \mbox{e}^{-q_{t}\tau_{t}}\Phi(d_{t}).
\end{equation*}


The Leland delta, as proposed by \cite{Leland}, presents a variation of the option replication approach outlined in the work of \cite{BlackScholes}, incorporating parameters such as the transaction cost proportion $\kappa$ and the rebalancing frequency $\lambda$.

\begin{equation*}
    \Delta_{t}^{L}= \mbox{e}^{-q_{t}\tau_{t}}\Phi\left(\tilde{d}_{t} \right),
\end{equation*}

where $\tilde{d}_{t}=\frac{\mbox{log}\left( \frac{S_{t}}{K} \right)+\left( r_{t}-q_{t}+\frac{1}{2}\tilde{\sigma}_{t}^2 \right)\tau_{t}}{\tilde{\sigma}_{t} \sqrt{\tau_{t}}}$ with $\tilde{\sigma}_{t}^{2} = \sigma (M_{t},\tau_{t},\beta_{t})^{2}\left[ 1+\sqrt{2/\pi}\frac{2\kappa}{\sigma (M_{t},\tau_{t},\beta_{t})\sqrt{\lambda}}\right]$.


\cite{Frnacois2022} provide the delta associated to the IV surface under model \eqref{full_5factormodel}:
\begin{equation*}
    \Delta_{t}^{SI} = e^{-q_{t}\tau_{t}} \left( \Phi(\mathcal{d}_{t,1})+\phi(\mathcal{d}_{t,1})\frac{\partial \sigma}{\partial M} \right)
\end{equation*}
with
$$\mathcal{d}_{t,1}=\frac{M_{t}}{\sigma(M_{t},\tau_{t},\beta_{t})}+\frac{1}{2}\sigma(M_{t},\tau_{t},\beta_{t})\sqrt{\tau_{t}},$$
and
\begin{align*}
     \frac{\partial \sigma}{\partial M} &=\beta_{t,3}\mathbbm{1}_{\{M_{t}\geq 0\}}+\beta_{t,3}\left(1-  \left(\frac{e^{2M_{t}}-1}{e^{2M_{t}}+1}\right)^{2} \right)\mathbbm{1}_{\{M_{t}< 0\}}+\beta_{t,4}2M_{t}e^{-M_{t}^{2}}\mbox{log}\left(\frac{T}{T_{max}} \right)\\
     &-\beta_{t,5}81M_{t}^{2}e^{27M_{t}^{3}}\mbox{log}\left(\frac{T}{T_{max}} \right)\mathbbm{1}_{\{M_{t}< 0\}}.
 \end{align*}

\pagebreak
\centerline{\Large \textbf{Supplementary material (not part of the paper)}}

\section{Details for the MSGD training approach}\label{appen:MSGDtraining}

The MSGD method estimates the penalty function $\mathcal{O}(\theta)$, which is typically unknown, through small samples of the hedging error called batches. Let $\mathbb{B}_{j}=\{ \xi_{T,i}^{\tilde{\delta}_{\theta_{j}}} \}_{i=1}^{N_{\mbox{\scriptsize{batch}}}}$ be the $j$-th batch where $N_{\mbox{\scriptsize{batch}}}$ is the batch size and $\xi_{T,i}^{\tilde{\delta}_{\theta_{j}}}$ denotes the hedging error of the $i$-th path in the $j$-th batch defined as
$$\xi_{T,i}^{\tilde{\delta}_{\theta_{j}}} = \Psi(S_{T,(j-1)N_{\mbox{\scriptsize{batch}}}+i}) - V_{T,i}^{\tilde{\delta}_{\theta_{j}}} \quad \mbox{for} \quad i\in\{1,\dots,N_{\mbox{\scriptsize{batch}}}\}, \, j\in\{1,...N \},$$
where $S_{T,(j-1)N_{\mbox{\scriptsize{batch}}}+i}$ is the price of the underlying asset at time $T$ in the ($(j-1)N_{\mbox{\scriptsize{batch}}}+i$)-th simulated path, $V_{T,i}^{\tilde{\delta}_{\theta_{j}}}$ is the terminal value of the hedging strategy for that path when $\theta = \theta_{j}$ and the simulated states are $X_{i}$. 

The penalty function estimation for the batch $\mathbb{B}$ is
\begin{align*}
    \Hat{C}^{\left(\mbox{\scriptsize{MSE}}\right)}(\theta_{j},\mathbb{B}_{j})&=\frac{1}{N_{\mbox{\scriptsize{batch}}}}\sum_{i=1}^{N_{\mbox{\scriptsize{batch}}}}\left(\xi_{T,i}^{\tilde{\delta}_{\theta_{j}}}\right)^{2}\, \mbox{,}\\
    \Hat{C}^{\left(\mbox{\scriptsize{SMSE}}\right)}(\theta_{j},\mathbb{B}_{j})&=\frac{1}{N_{\mbox{\scriptsize{batch}}}}\sum_{i=1}^{N_{\mbox{\scriptsize{batch}}}}\left(\xi_{T,i}^{\tilde{\delta}_{\theta_{j}}}\right)^{2}\mathbbm{1}_{\left\{\xi_{T,i}^{\tilde{\delta}_{\theta_{j}}}\geq 0\right\}}\, \mbox{,}\\
    \Hat{C}^{\left(\mbox{\scriptsize{CVaR}}\right)}(\theta_{j},\mathbb{B}_{j})&= \widehat{\mbox{VaR}}_{\alpha}(\mathbb{B}_{j})+\frac{1}{(1-\alpha)N_{\mbox{\scriptsize{batch}}}}\sum_{i=1}^{N_{\mbox{\scriptsize{batch}}}}\max\left( \xi_{T,i}^{\tilde{\delta}_{\theta_{j}}}-\widehat{\mbox{VaR}}_{\alpha}(\mathbb{B}_{j}),0 \right)\, \mbox{,}
\end{align*}

where $\widehat{\mbox{VaR}}_{\alpha}(\mathbb{B}_{j})=\xi_{T,\lceil \alpha \cdot N_{\mbox{\scriptsize{batch}}} \rceil}^{\tilde{\delta}_{\theta_{j}}}$ is the estimation of the VaR obtained from the ordered sample $\{ \xi_{T,[i]}^{\tilde{\delta}_{\theta_{j}}} \}_{i=1}^{N_{\mbox{\scriptsize{batch}}}}$ and $\lceil \cdot \rceil$ is the ceiling function. These empirical approximations are used to estimate the gradient of the penalty function required in Equation \eqref{updatingrule}.\footnote{In particular, the gradient of these estimations has analytical expressions for FFNN, LSTM networks and thus for RNN-FNN networks. Details about gradient of the empirical objective function are provided in \cite{goodfellow2016deep}.} 

The selection of batch size plays a key role in the MSGD training approach as we empirically measure tail risk. A larger batch size provides a more accurate gradient estimate by averaging gradients computed over more simulated paths, thereby promoting stable convergence during training. However, large batch sizes may introduce certain disadvantages such as increased memory requirements, slower convergence, and potential generalization issues. We adopt the batch size used in \cite{carbonneau2021deep} ($N_{\mbox{\scriptsize{batch}}}=1,\!000$), which achieves a good balance between accuracy and convergence.

\section{Joint Implied Volatility and Return model}\label{appen:JIVR}

\subsection{Daily implied volatility surface}\label{subappen:IVsurface}

The functional representation of the IV surface model introduced by \cite{Frnacois2022} is
\begin{align}
     & \sigma (M_{t},\tau_{t},\beta_{t}) =  \underbrace{\beta_{t,1}}_{f_{1}\mbox{: \scriptsize{Long-term ATM IV}}} +  \beta_{t,2}\underbrace{e^{-\sqrt{\tau_{t} /T_{conv}}}}_{f_{2}\mbox{: \scriptsize{Time-to-maturity slope}}} + \beta_{t,3}\underbrace{\left( M_{t}\mathbbm{1}_{\{ M_{t}\geq 0 \}} + \frac{e^{2M_{t}}-1}{e^{2M_{t}}+1}\mathbbm{1}_{\{M_{t}< 0\}} \right)}_{f_{3}\mbox{: \scriptsize{Moneyness slope}}} \nonumber\\
     & + \beta_{t,4}\underbrace{\left( 1-e^{-M_{t}^{2}} \right)\mbox{log}(\tau_{t}/T_{max})}_{f_{4}\mbox{: \scriptsize{Smile attenuation}}}+\beta_{t,5}\underbrace{\left( 1-e^{(3M_{t})^{3}} \right)\mbox{log}(\tau_{t}/T_{max})\mathbbm{1}_{\{M_{t}< 0\}}}_{f_{5}\mbox{: \scriptsize{Smirk}}}\, , \quad \tau_{t} \in [T_{min},T_{max}] 
     \label{full_5factormodel}
\end{align}

where $T_{max}$ is set to 5 years, $T_{min}=6/252$ and $T_{conv}=0.25$. As in \cite{dumas1998implied},  a minimum threshold of 0.01 is applied to the volatility surface to prevent negative values.

\subsection{Joint Implied Volatility and Return}\label{subappen:JIVR_timeseries}

The JIVR model proposed by \cite{Frnacois2023} has 6 components, one for the underlying asset excess returns and the other 5 for fluctuations of the IV surface coefficients. The S\&P 500 excess return follows:
 \begin{align}
     R_{t+1} & = \xi_{t+1}-\psi(\sqrt{h_{t+1,R}\Delta}) + \sqrt{h_{t+1,R}\Delta}\epsilon_{t+1,R}, \nonumber\\
     h_{t+1,R} & = Y_{t}+\kappa_{R}(h_{t,R}-Y_{t})+a_{R}h_{t,R}(\epsilon_{t,R}^{2}-1-2\gamma_{R}\epsilon_{t,R}), \label{return}\\
     Y_{t} &= \left(\omega_{R}\, \sigma \left(0,\frac{1}{12},\beta_{t}\right)\right)^2, \nonumber
     \label{returnseries}
 \end{align}
where the equity risk premium is
\begin{equation}
    \xi_{t+1}=\psi(-\lambda\sqrt{h_{t+1,R}\Delta})- \psi((1-\lambda)\sqrt{h_{t+1,R}\Delta}) + \psi(\sqrt{h_{t+1,R}\Delta}),
\end{equation}
the process $\{\epsilon_{t,R}\}^T_{t=0}$ is a sequence of iid standardized NIG random variables with parameters $\zeta_{R}$ and $\varphi_{R}$,\footnote{The standard NIG random variable $\epsilon$ follows the probability density function with parameters $\zeta$ and $\varphi$: $$f(x)=\frac{B_{1}\left( \sqrt{\frac{\varphi^{6}}{\varphi^{2}+\zeta^{2}}+(\varphi^{2}+\zeta^{2})\left(x+ \frac{\varphi^{2}\zeta}{\varphi^{2}+\zeta^{2}} \right)^{2}} \right)}{\pi\sqrt{\frac{1}{\varphi^{2}+\zeta^{2}}+\frac{\varphi^{2}+\zeta^{2}}{\varphi^{6}}\left(x+ \frac{\varphi^{2}\zeta}{\varphi^{2}+\zeta^{2}} \right)^{2}} } e^{\left( \frac{\varphi^{4}}{\varphi^{2}+\zeta^{2}}+\zeta\left(x+ \frac{\varphi^{2}\zeta}{\varphi^{2}+\zeta^{2}} \right) \right),}$$ where $B_{1}(\cdot)$ denotes the modified Bessel function of the second kind with index 1. The common $(\alpha, \beta, \delta, \mu)$-specification can be obtained by replacing $\beta$ and $\gamma$ ($\gamma = \sqrt{\alpha^2-\beta^2}$), with $\zeta$ and $\varphi$, respectively, and imposing a null mean and unit variance to express $\delta$ and $\mu$ interms of $\alpha$, $\beta$.} 
and $\psi$ is their cumulant generating function.\footnote{For $-\sqrt{\zeta^{2}+\varphi^{2}}-\zeta<z<\sqrt{\zeta^{2}+\varphi^{2}}-\zeta$, the cumulant generating function is given by 
$$\psi(z)=\frac{\varphi^{2}}{\varphi^{2}+\zeta^{2}}\left( -\zeta z+\varphi^{2}-\varphi\sqrt{\varphi^{2}+\zeta^{2}-(\varphi+\zeta)^{2}} \right).$$}
Parameters of the excess return component of the model are thus $\Theta_{R} = (\lambda,\kappa_{R},\gamma_{R},a_{R},\omega_{R},\zeta_{R},\varphi_{R})$. 

The evolution of the long-term factor $(\beta_{1})$ is
\begin{align}
     \beta_{t+1,1}&= \alpha_{1}+\sum_{i=1}^{5}\theta_{1,j}\beta_{t,j}+\sqrt{h_{t+1,1}\Delta}\epsilon_{t+1,1}, \nonumber\\
     h_{t+1,1} &= U_{t}+\kappa_{1}(h_{t,1}-U_{t})+a_{1}h_{t,1}(\epsilon_{t,1}^{2}-1-2\gamma_{1}\epsilon_{t,1}), \label{beta1series}\\
     U_{t} &= \left(\omega_{1}\cdot \sigma\left(0,\frac{1}{12},\beta_{t}\right)\right)^2. \nonumber
 \end{align}

For the other 4 coefficients, $i\in\{2,3,4,5 \}$, the time evolution satisfies
\begin{align}
    \beta_{t+1,i} &= \alpha_{i}+\sum_{j=1}^{5}\theta_{i,j}\beta_{t,j}+\nu\beta_{t-1,2}\mathds{1}_{\{i=2\}}+\sqrt{h_{t+1,i}\Delta}\epsilon_{t+1,i}, \nonumber\\
    h_{t+1,i} &= \sigma_{i}^{2}+\kappa_{i}(h_{t,i}-\sigma_{i}^{2})+a_{i}h_{t,i}(\epsilon_{t,i}^{2}-1-2\gamma_{i}\epsilon_{t,i}),
    \label{restofbetas}
\end{align}

where $\{\epsilon_{t,i} \}_{i=1}^{5}$ are time-independent standardized NIG random variables with parameters $\{ (\zeta_{i},\varphi_{i}) \}_{i=1}^{5}$, respectively. Parameters are $\{\omega_{1},\nu, \Theta_{1}, \Theta_{2}, \Theta_{3}, \Theta_{4}, \Theta_{5}\}$ with $\Theta_{i} = (\alpha_{i},\{\theta_{i,1}\}_{i=1}^{5},\sigma_{i},\kappa_{i},a_{i},\gamma_{i},\zeta_{i},\varphi_{i})\}_{i=1}^{5}$.

The JIVR model also imposes a dependence structure on contemporaneous innovations $\epsilon_{t} = (\epsilon_{t,R},\epsilon_{t,1},...,\epsilon_{t,5})$ through a Gaussian copula parameterized in terms of a covariance matrix $\Sigma$ of dimension ${6\times 6}$. 

Estimates of all JIVR model parameters are taken from Table 5 and Table 6 of \cite{Frnacois2023}.

\section{JIVR Model parameters}

\begin{table}[H]
\centering
\renewcommand{\arraystretch}{1.5}
\caption{Estimated Gaussian copula parameters}
\begin{tabular}{lrrrrrr}
\toprule
 & $\epsilon_{t,R}$ & $\epsilon_{t,1}$ & $\epsilon_{t,2}$ & $\epsilon_{t,3}$ & $\epsilon_{t,4}$ & $\epsilon_{t,5}$ \\
\midrule
$\epsilon_{t,R}$ & 1.000 &  &  &  & &  \\
$\epsilon_{t,1}$ & -0.550 & 1.000 &  &  &  &  \\
$\epsilon_{t,2}$ & -0.690 & 0.140 & 1.000 &  & &  \\
$\epsilon_{t,3}$ & 0.030 & -0.030 & -0.0100 & 1.000 &  &  \\
$\epsilon_{t,4}$ & -0.220 & 0.250 & 0.120 & 0.280 & 1.000 &  \\
$\epsilon_{t,5}$ & -0.340 & 0.170 & 0.370 & 0.130 & -0.050 & 1.000 \\
\bottomrule
\end{tabular}
\end{table}

\begin{table}[H]
\centering
\renewcommand{\arraystretch}{1.5}
\caption{ JIVR model parameter estimates }
\begin{tabular}{lrrrrrrr}
\toprule
Parameter  & $\beta_1$ & $\beta_2$ & $\beta_3$ & $\beta_4$ & $\beta_5$ & & S\&P500 \\
\cline{1-6}\cline{8-8}
$\alpha$ & 0.000899 & 0.008400 & 0.000770 & -0.001393 & 0.000657 & $\lambda$ & 2.711279 \\
$\theta_1$ & 0.996290 & -0.013869 &  & 0.002841 &  &  &\\
$\theta_2$ & 0.003669 & 0.877813 & 0.001300 &  &  &  &\\
$\theta_3$ &  & -0.032640 & 0.997071 & 0.003722 & -0.004198 &  &\\
$\theta_4$ &  &  &  & 0.980269 &  &  &\\
$\theta_5$ &  & -0.047789 &  &  & 0.986019 &  &\\
$\nu$ &  & 0.089445 &  &  &  &  &\\
$\sigma\sqrt{252}$ &  & 0.380279 & 0.052198 & 0.048641 & 0.051536 &  &\\
$\omega$ & 0.267589 &  &  &  &  & & 0.977291 \\
$\kappa$ & 0.838220 & 0.965751 & 0.974251 & 0.945377 & 0.980844 & & 0.888977 \\
$a$ & 0.134152 & 0.098272 & 0.092646 & 0.102201 & 0.100502 & & 0.056087 \\
$\gamma$ & -0.111813 & -1.482862 & 0.096766 & 0.060558 & -0.102996 & & 2.507796 \\
$\zeta$ & 0.143760 & 0.852943 & 0.029109 & -0.159051 & 0.092664 & & -0.641306 \\
$\varphi$ & 1.351070 & 1.538928 & 2.284780 & 1.449977 & 1.428477 & & 2.039669 \\
\bottomrule
\end{tabular}
\end{table}


\section{Network fine-tuning}

\subsection{Bounded strategies - leverage constraint}\label{se:Bounded strategies - leverage constraint}

This section presents a comparison of the hedging performance of RL agents trained under the full state space considering the CVaR$_{95\%}$ as penalty function. The comparison is conducted considering two RL agents: (i) RL-CVaR$_{95\%}$-LC, agent subject to a leverage constraint  equivalent to the initial value of the underlying asset, set at $B=100$, and (ii) RL-CVaR$_{95\%}$, agent operating without leverage constraints. This comparison evaluates the estimated values of all penalty functions and the average Profit and Loss (Avg P\&L). This analysis aims to elucidate the impact of boundary conditions on RL agent behavior and performance when hedging of a European ATM call option with a maturity of $N=63$ days in the absence of transaction costs, i.e., $\kappa = 0\%$).

\autoref{table:cash_constraint} outlines that the agent without a leverage constraint shows more profitable strategies; however, it also leads to very large losses. For instance, out-of-sample CVaR$_{99\%}$ and SMSE metrics exhibit higher values for this agent, compared to its counterpart with a leverage constraint. 

\begin{table}[H]
\centering
\renewcommand{\arraystretch}{1.5}
\caption{RNN-FNN hedging error statistics of a short position in a ATM call option with maturity of 63 days.}
\begin{tabular}{p{2.5cm} >{\centering\arraybackslash}p{3.2cm} >{\centering\arraybackslash}p{3.2cm}}
\hline
Function & RL-CVaR$_{95\%}$-LC & RL-CVaR$_{95\%}$ \\
\hline
Avg P\&L & 0.440 &   {\bfseries 0.715} \\
CVaR$_{95\%}$ & 1.294 &   {\bfseries 1.242} \\
CVaR$_{99\%}$ & {\bfseries 2.502} &   2.622 \\
MSE & {\bfseries 1.205} &   14.650 \\
SMSE & {\bfseries 0.241} &   0.327 \\
\hline
\end{tabular}
\label{table:cash_constraint}
\begin{tablenotes}
\item \footnotesize Results are computed using 100,000 out-of-sample paths in the absence of transaction costs ($\kappa = 0\%$). Agents are trained under the full state space $(V_{t}^{\delta},\delta_{t},\tau_{t},S_{t},\{\beta_{t,i} \}^5_{i=1}, h_{t,R}, \{ h_{t,i}\}^5_{i=1})$ according to the conditions outlined in Section \ref{subsub:network_architecture}. The average option price is \$3.89 with a standard deviation of \$1.29. RL-CVaR$_{95\%}$-LC denotes the RL agent trained with a leverage constraint, while RL-CVaR$_{95\%}$ refers to the agent without it. Best performances are highlighted in bold. 
\end{tablenotes}
\end{table}

\begin{figure}[h]\centering
\caption{Doubling strategy dynamics for a short position in a ATM call option with a maturity of 63 days.}
\includegraphics[width=17cm]{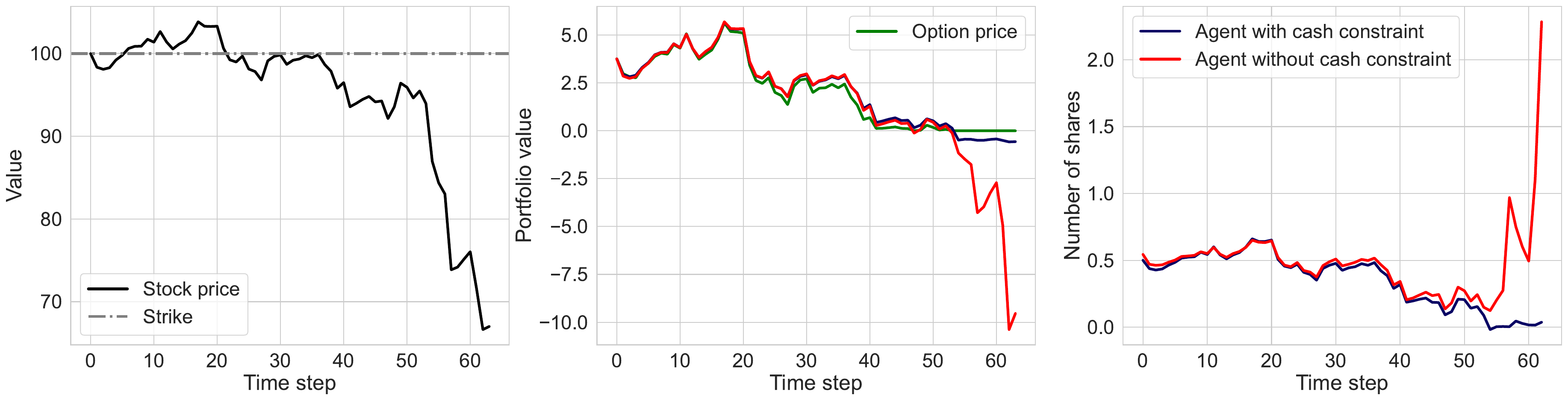}
\label{fig:Martingale_startegies}
\begin{tablenotes}
\item \footnotesize Results are obtained using a deep OTM path in the absence of transaction costs ($\kappa = 0\%$), with agents trained according to the conditions outlined in Section \ref{subsub:network_architecture}. The training encompasses the full state space and utilizes CVaR$_{95\%}$ as a penalty function. The agent referred to as "Agent with cash constraint" incorporates a leverage constraint of $B=100$, whereas the agent labeled "Agent without cash constraint" does not.
\end{tablenotes}
\end{figure}

Moreover, our numerical experiments demonstrate that the agent without leverage constraints learns doubling strategies, which are incompatible with sound risk management practices. For instance, \autoref{fig:Martingale_startegies} illustrates such behavior through three panels associated with the hedging process of a deep OTM path (first panel).\footnote{For the purpose of this experiment, a deep OTM path refers to the trajectory of the underlying asset that keeps the option significantly out-of-the-money.} We observe that the agent, without a leverage constraint, tends to increase its position in the underlying (third panel) when a loss in the portfolio value is observed (second panel), aiming to recover the loss over the long run with doubling strategies. Conversely, agents trained with a leverage constraint control their position in the underlying asset, resulting in the learning of different and less risky strategies. 

\subsection{Network architecture selection}\label{appen:architecture_selection}

In this section, we investigate the superiority of the RNN-FNN architecture compared to conventional architectures introduced in deep hedging literature, such as the FFNN and the LSTM architectures. In line with our previous experiments, we consider four penalty functions to train the agents under the full state space: CVaR$_{95\%}$, CVaR$_{99\%}$, MSE, and SMSE. Again, our experiment focuses on hedging a short position of a European ATM call option with a maturity of $N=63$ days, assuming no transaction costs.

\begin{figure}[h]\centering
\caption{Network performance for a short position in a ATM call option with a maturity of 63 days.}
\includegraphics[width=16cm]{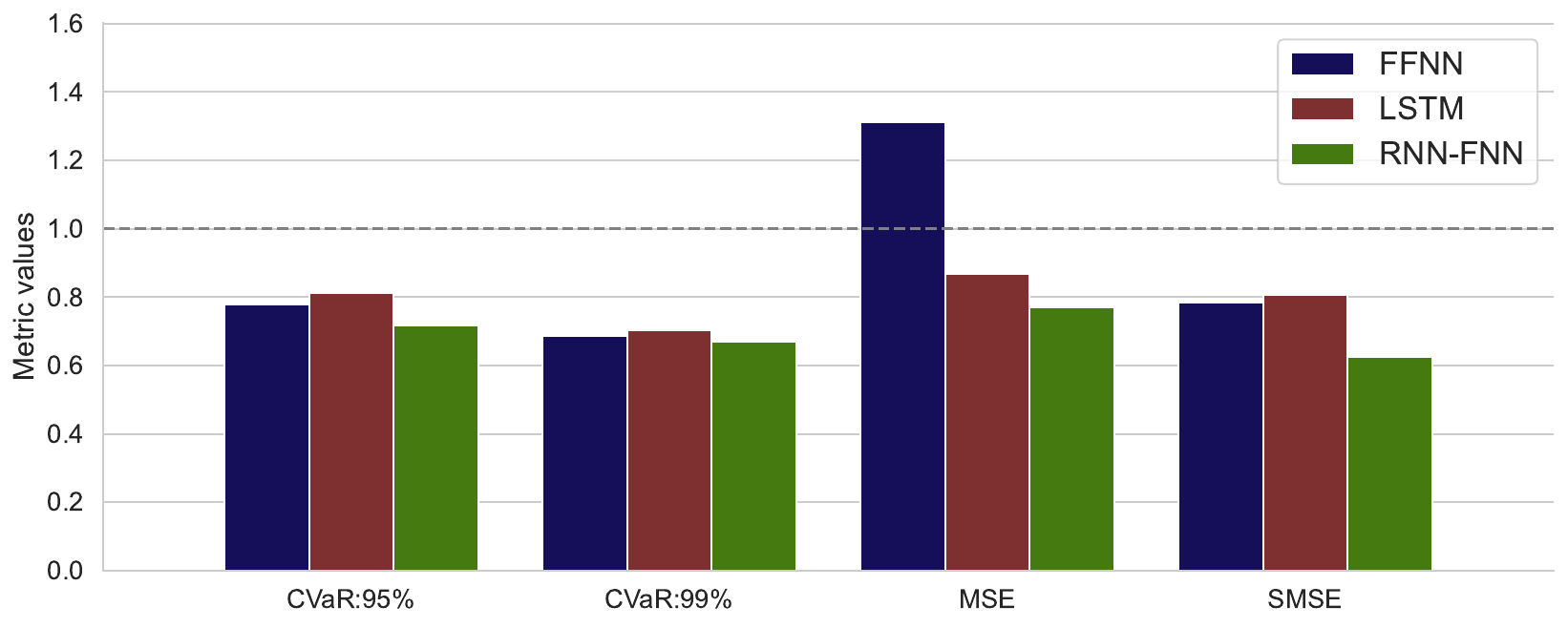}
\begin{tablenotes}
\item 
\footnotesize Results are computed using 100,000 out-of-sample paths in the absence of transaction costs ($\kappa = 0\%$). Agents are trained under the reduced state space $(\tau_{t},S_{t},\{\beta_{t,i} \}^5_{i=1}, h_{t,R}, \{ h_{t,i}\}^5_{i=1})$ according to the conditions outlined in Section \ref{subsub:network_architecture}. The setup for the different networks follows the architecture described in Section \ref{se:NNARCHITECTURE}, with $L_{1}=0$ and $L_{2}=4$ for the FFNN, $L_{1}=4$ and $L_{2}=0$ for the LSTM, and $L_{1}=2$ and $L_{2}=2$ for RNN-FNN. Results show optimal values obtained from agents trained under CVaR$_{95\%}$, CVaR$_{99\%}$, MSE, and SMSE for the three networks. These values are normalized by the estimated values of each penalty function obtained with DH.
\end{tablenotes}
\label{fig:networks_performance}
\end{figure}

The superiority of the RNN-FNN over the LSTM and the FFNN is evaluated based on the optimal value of each penalty function. \autoref{fig:networks_performance} illustrates the optimal values of each penalty function for each architecture, normalized by the estimated value obtained with DH. Notably, the RNN-FNN setup considered in this paper significantly outperforms both the benchmark and other architectures. Conversely, the FFNN does not surpass the benchmark for the MSE, and the LSTM exhibits almost the same performance as the benchmark.

\begin{figure}[h]\centering
\caption{Neural network performance for a short position in an ATM call option with a maturity of 63 days: sensitivity to the state of the economy.}
\includegraphics[width=16cm]{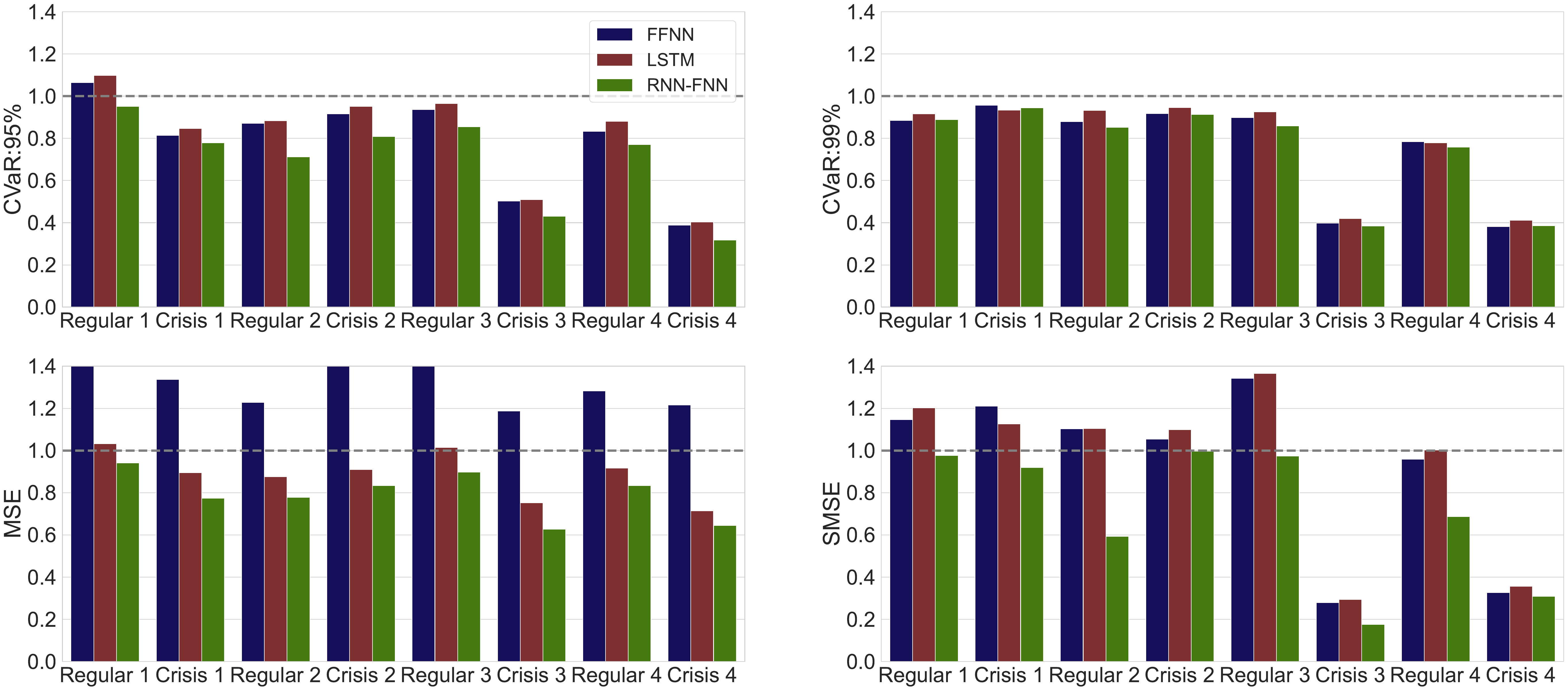}
\begin{tablenotes}
\item \footnotesize Results are computed using 100,000 out-of-sample paths in the absence of transaction costs ($\kappa = 0\%$). Agents are trained under the reduced state space $(\tau_{t},S_{t},\{\beta_{t,i} \}^5_{i=1}, h_{t,R}, \{ h_{t,i}\}^5_{i=1})$ according to the conditions outlined in Section \ref{subsub:network_architecture}. Each panel illustrates the values obtained by different architectures for the following hedging metrics: Avg P\&L, CVaR$_{95\%}$, MSE, and SMSE.  These metrics are normalized by the estimated values of each penalty function obtained with DH. The setup for the different networks follows the architecture described in Section \ref{se:NNARCHITECTURE}, with $L_{1}=0$ and $L_{2}=4$ for the FFNN, $L_{1}=4$ and $L_{2}=0$ for the LSTM, and $L_{1}=2$ and $L_{2}=2$ for RNN-FNN.\footnotemark
\label{fig:networks_state_economy}
\end{tablenotes}
\end{figure}

A second test to demonstrate the superiority of our architecture involves computing the optimal values of the penalty functions over various clusters of paths. The objective is to isolate the impact of different state of the economy on performance and assess the robustness of our architecture. \autoref{fig:networks_state_economy} displays the optimal values of each architecture normalized by the estimated values obtained by DH for all penalty functions. Notably, RNN-FNN agents outperform both the benchmark and other architectures across all penalty functions, regardless of the economic conditions under which the simulations were conducted. Conversely, FFNN and LSTM agents fail to outperform the benchmark across all states of the economy when the agents are trained under the CVaR$_{95\%}$, MSE and SMSE penalty functions, as shown in the top-left, bottom-left and bottom-right panels, respectively.

Results obtained under various economic conditions unequivocally demonstrate the superiority of the RNN-FNN network across all performance metrics. Moreover, the RNN-FNN not only offers better performance in terms of risk management but also reduces computational costs and implementation complexity, with an average reduction in training time of 46\%.

\footnotetext{\label{fn:periods} The periods aim to approximate different states of the economy considering the time frames specified in \autoref{tab:time_frames}.}


\subsection{Dropout parameter selection}\label{appen:dropout_parameter}

The process of selecting the dropout parameter for the regularization method involved evaluating the performance of four agents across a range of potential parameter values. These agents are trained using the full state space, considering four penalty functions: CVaR$_{95\%}$, CVaR$_{99\%}$, MSE, and SMSE. The performance of the agents is measured in terms of the estimated values of the penalty functions after hedging a short position of a European call option with a maturity of $N=63$ days, with no transaction costs.

\begin{figure}[h]\centering
\caption{RNN-FNN performance for a short position in an ATM call option with a maturity of 63 days: the effect of the dropout parameter in the training phase.}
\includegraphics[width=16cm]{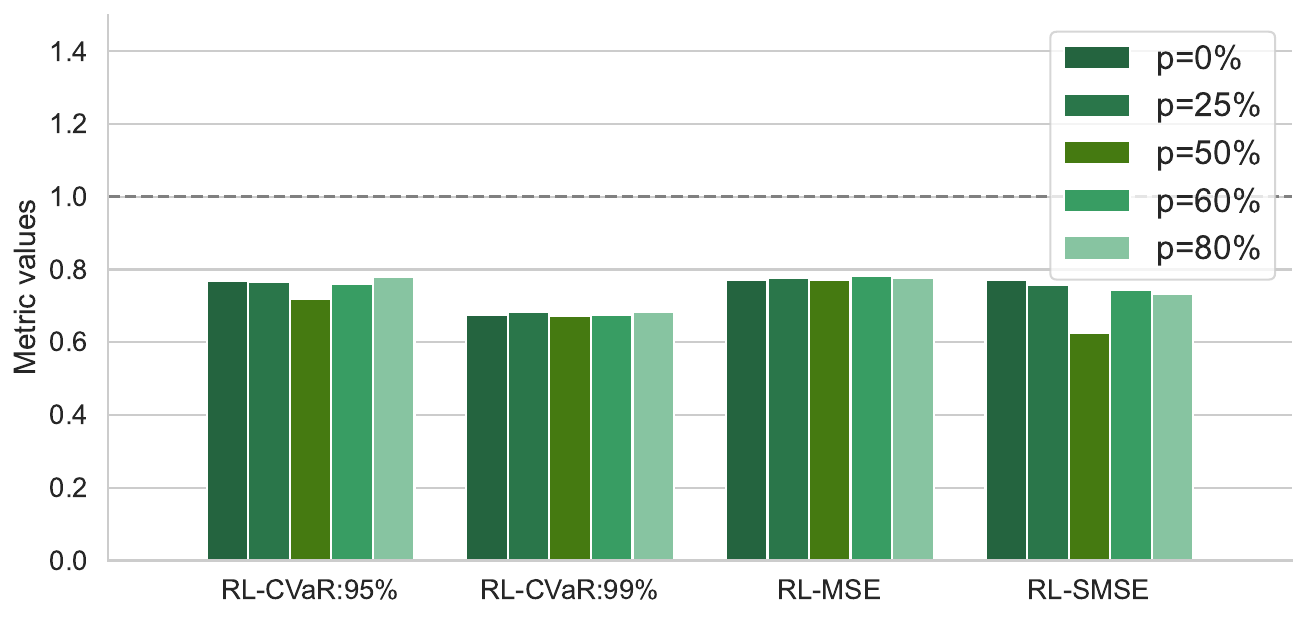}
\begin{tablenotes}
\item  \footnotesize Results are computed using 100,000 out-of-sample paths in the absence of transaction costs ($\kappa = 0\%$). Agents are trained under the reduced state space $(\tau_{t},S_{t},\{\beta_{t,i} \}^5_{i=1}, h_{t,R}, \{ h_{t,i}\}^5_{i=1})$ according to the conditions outlined in Section \ref{subsub:network_architecture}. All the metrics are expressed in proportion of values obtained under DH.
\end{tablenotes}
\label{fig:dropout_param}
\end{figure}

\autoref{fig:dropout_param} illustrates a comparison of the optimal values of penalty functions normalized by the estimated values obtained using DH. It is noteworthy that the RNN-FNN network consistently outperforms the benchmark regardless of the dropout parameter value. Additionally, the selection of the dropout parameter remains consistent across all objective functions, with a dropout probability ($p$) of 50\% yielding optimal performance. Consequently, for all our experiments, a dropout regularization parameter of $p=50\%$ is adopted. This parameter drives the likelihood of randomly dropping out a fraction of the units within a neural network during training, thereby generating varied architectures for each training iteration. As demonstrated in \cite{warde2013empirical}, this method not only effectively mitigates overfitting but also boosts performance, which is consistent with our numerical results.


\section{Quadratic hedging problem}\label{appen:quadratic_problem}

This appendix aims to compare the closed-form solution of the quadratic hedging problem as outlined by \cite{godin2019closed} with our own approach within the framework of Black-Scholes market dynamics, where log-returns are assumed to adhere to a Gaussian distribution. \autoref{tab:QH_problem} illustrates performance metrics using three distinct penalty functions from two experiments analyzing the hedging error of an ATM call option with a strike price of K=100, with maturities of 63 days and 252 days. The experiments involve 16 time steps and 5 time steps for rebalancing, respectively. 

The outcomes of the 63-day maturity ATM option reveal that the RL agents outperform the two benchmarks, Black-Scholes delta (BS) and the quadratic hedging solution (QH), examined in this study. Additionally, the performance gap between the RL agent trained with the full state space (RL-Full) and the agent trained with the reduced state space (RL-Reduced) is negligible. In fact, the Kolmogorov-Smirnov test fails to reject the null hypothesis that the hedging errors of both agents are equally distributed, with a confidence level of 99.9\%.\footnote{The Kolmogorov-Smirnov test, outlined in \cite{KStest}, is a non-parametric statistical test used to determine whether two samples differ significantly.}

\begin{table}[H]
\centering
\renewcommand{\arraystretch}{1.5}
\caption{RNN-FNN hedging error statistics for a short position ATM call option with two different maturities and rebalancing periods under the MSE as penalty function.}
\begin{tabular}{p{1.7cm} >{\centering\arraybackslash}p{1.3cm} >{\centering\arraybackslash}p{1.3cm} >{\centering\arraybackslash}p{1.3cm} >{\centering\arraybackslash}p{1.3cm} c >{\centering\arraybackslash}p{1.3cm}>{\centering\arraybackslash}p{1.3cm}>{\centering\arraybackslash}p{1.3cm}>{\centering\arraybackslash}p{1.3cm}}
\hline
 \multicolumn{1}{c}{} & \multicolumn{4}{c}{Maturity: 63, Time steps:16} & & \multicolumn{4}{c}{Maturity: 252, Time steps:5} \\

\cline{2-5}\cline{7-10}

Function & BS & QH & RL-F & RL-R & & BS & QH & RL-F & RL-R \\
\hline
Avg P\&L & 0.005 & -0.081 & -0.014 & -0.009 &   & 0.185 & -0.335 & -0.279 & -0.216\\
CVaR$_{95\%}$ & 1.942 & 2.619 & 1.897 & 1.931 &   & 7.291 & 6.662 & 6.596 & 6.514 \\
CVaR$_{99\%}$ & 2.896 & 3.748 & 2.808 & 2.881 &   & 10.62 & 10.488 & 9.913 & 9.624 \\
MSE & 0.684 & 1.272 & 0.681 & 0.683 &   & 8.691 & 7.609 & 7.865 & 7.834 \\
SMSE & 0.367 & 0.647 & 0.350 & 0.359 &   & 5.292 & 3.721 & 3.852 & 3.941 \\
\hline
\end{tabular}
\begin{tablenotes}
\item \footnotesize These results are computed considering the hedging error of 99,000 out-of-sample independent paths from the Black-Scholes market with yearly parameters $\mu = 0.0892$ and and $\sigma = 0.1952$. The RNN-FNN is trained based on 400,000 independent paths under the same scheme.
\end{tablenotes}
\label{tab:QH_problem}
\end{table}

Conversely, results of the 252-day maturity ATM option reveal that the QH approach exhibits slightly superior performance in MSE, as anticipated due to its closed-form nature. However, the performance of RL agents demonstrates stronger potential in terms of risk management, evident from their ability to yield the lowest CVaR values and closely approximate the MSE of the closed-form solution. Moreover, the agent trained with the reduced state space (RL-Reduced) exhibits enhanced performance compared to its counterpart trained with the full state space. 

\begin{figure}[h]\centering
\caption{RNN-FNN loss function for a short position ATM call option with maturity $N=63$ days and 16 time steps for rebalancing.}
\includegraphics[width=16cm]{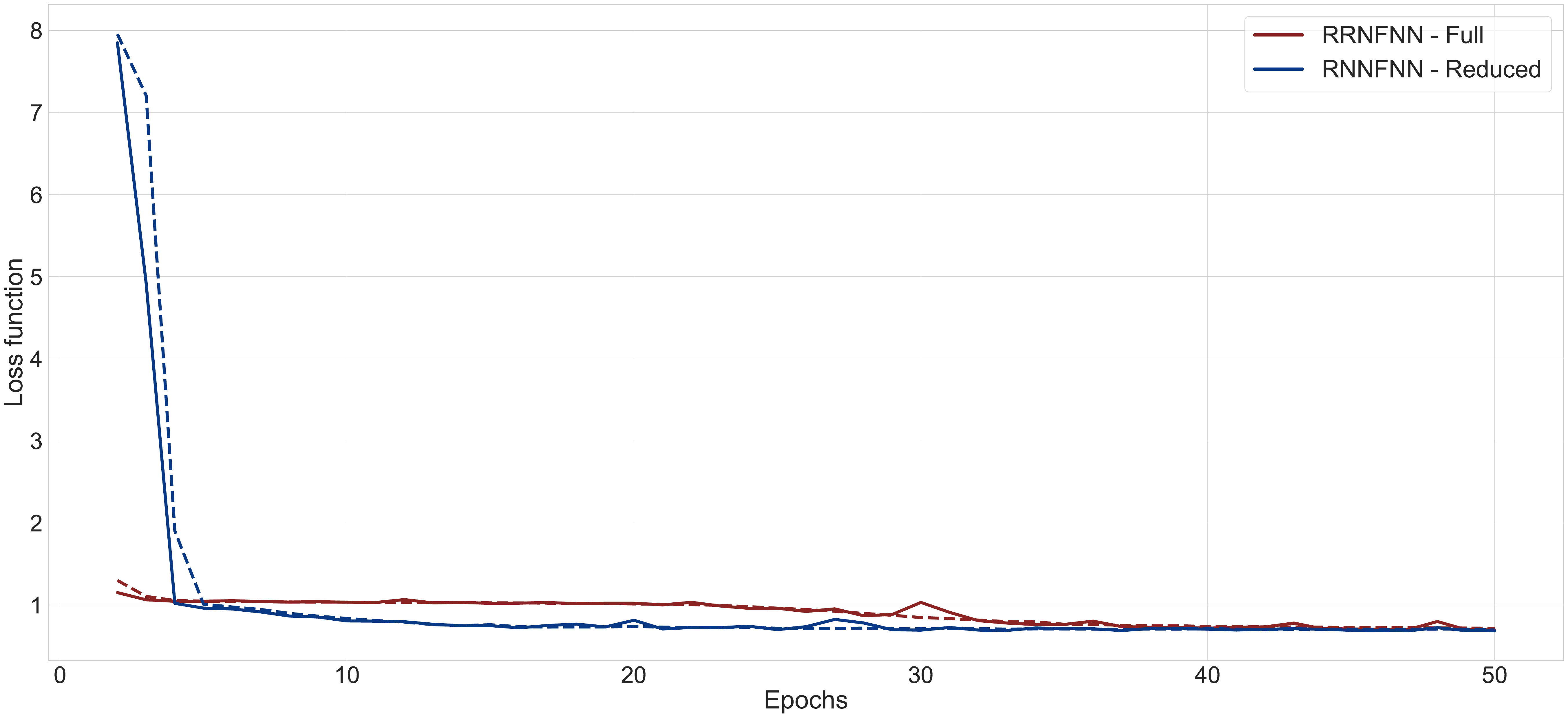}
\begin{tablenotes}
\item \footnotesize Results are computed considering the MSE as the loss functions and the hedging error of 99,000 out-of-sample independent paths with random initialization for the validation loss curve (full line) and 400,000 independent paths fot the training loss curve (dash line). Simulations are performed based on the Black-Scholes model and agents are trained considering the cash constraint of $B=100$.
\end{tablenotes}
\label{fig:loss_curves_quadratic}
\end{figure}

Consistent with the findings detailed in Section \ref{subsub:state_space} regarding the dynamics of the JIVR model, the RL agent trained with the reduced state space exhibits improved the rate of convergence during the training phase. For instance, as depicted in \autoref{fig:loss_curves_quadratic}, the penalty curve evolution for the RL-MSE agent over 50 epochs illustrates this trend. This approach effectively reduces computational costs during training and accelerates convergence to optimal performance. These numerical outcomes confirm that our method achieves robust performance without necessitating the inclusion of portfolio value, even within the Quadratic Hedging framework with Black-Scholes market dynamics.


\section{In-sample backtest}

In this section, we benchmark our approach using a historical path of the JIVR model spanning from January 5, 1996, to December 31, 2020, to assess the effectiveness of RL agents. This experiment examines their performance based on the historical series $\{R_{t},\beta_{t}\}$. Specifically, we evaluate the hedging performance considering a new European ATM call option with a maturity of 63 days every 21 business days along this historical path. The option prices, serving as initial hedging portfolio values, are determined with the prevailing IV surface on the day the hedge is initiated.

\begin{figure}[H]\centering
\caption{Cumulative P\&L for ATM call options with a maturity of 63 days under real asset price dynamics.}
\includegraphics[width=17.5cm]{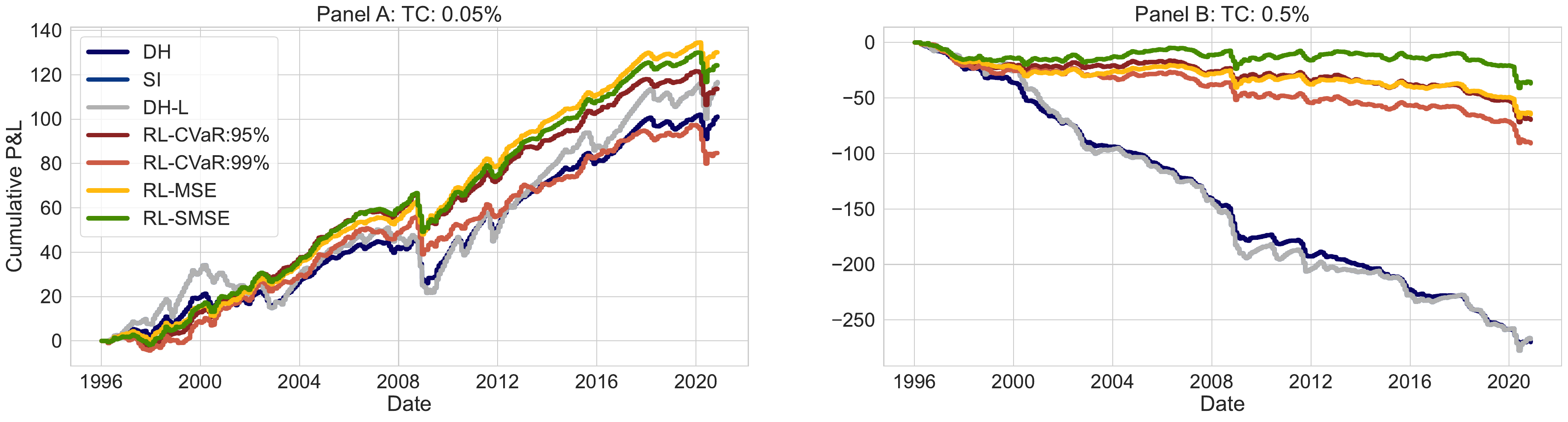}
\begin{tablenotes}
\item 
\footnotesize Results are computed based on the observed P\&L from hedging 296 simulated ATM Call options under real market conditions observed from May 1, 1996, to December 31, 2020. A new option is considered every 21 business days. Agents are trained under the reduced state space $(\delta_{t},\tau_{t},S_{t},\{\beta_{t,i} \}^5_{i=1}, h_{t,R})$ according to the conditions outlined in Section \ref{subsub:network_architecture}.
\end{tablenotes}
\label{fig:cumulative_p&l}
\end{figure}

To assess the robustness of the model under more general market conditions, we conduct a comparison of cumulative P\&Ls, which are computed as the cumulative sums of the P\&L achieved by each strategy at the maturity of each option during the analyzed period. As illustrated in \autoref{fig:cumulative_p&l}, which depicts the evolution of the cumulative P\&L across two panels, each representing a different transaction cost level, the gap in cumulative P\&L between RL agents and benchmarks significantly widens as transaction cost rates increase. This, again, highlights the adaptability of the RL approach to various market conditions. 

In contrast to the findings presented in Section \ref{subsub:benchmarking_general}, where SI delta yielded the highest profitability among all strategies, the RL approach achieved the highest profitability in historical paths. Specifically, agents trained under the MSE and SMSE emerged as the most profitable ones.

Moreover, RL agents consistently outperform benchmarks across most of the transaction cost levels. However, when the transaction cost is small (left panel), Benchmarks exhibit a better cumulative P\&L compared to the RL agent trained under the CVaR$_{99\%}$ penalty function at the end of the period, December 31, 2020. Nevertheless, benchmarks also demonstrate more variability and larger losses, such as those observed between 2008 and 2012. In contrast, RL agents show a smaller decrease in cumulative P\&L, indicating more resilience during crisis periods. In general, the observed performance of the RL agents under historical data is consistent with our findings under simulated data, indicating the robustness of our approach.

\end{document}